\documentclass[%
 reprint,
superscriptaddress,
 amsmath,amssymb,
 aps,
floatfix,
]{revtex4-2}

\newcommand{\topic}[1]{}

\newcommand{\aref}[1]{Appendix~\ref{#1}}
\newcommand{\eref}[1]{Eq.~(\ref{#1})}
\newcommand{\tref}[1]{Table.~\ref{#1}}
\newcommand{\fref}[1]{Fig.~\ref{#1}}
\newcommand{\fsref}[1]{Figs.~\ref{#1}}
\newcommand{\figureref}[1]{Figure~\ref{#1}}
\newcommand{\panel}[1]{(#1)}
\newcommand{\cpanel}[1]{\textbf{(#1)}}

\newcommand{\ketbra}[1]{\ket{#1}\bra{#1}}

\usepackage{graphicx}
\usepackage{dcolumn}
\usepackage{bm}
\usepackage[colorlinks]{hyperref}
\hypersetup{%
    plainpages=true,
    breaklinks=true,
    hypertexnames=false,
    pageanchor=true,
    colorlinks=true,
    linkcolor={blue},
    citecolor={blue},
    urlcolor={blue},
    anchorcolor={black}
}
\usepackage{hhline}

\usepackage{braket, siunitx}

\begin{document}

\preprint{APS/123-QED}

\title{High-Fidelity, Frequency-Flexible Two-Qubit Fluxonium Gates with~a~Transmon~Coupler}

\def\RLEaffil{Research Laboratory of Electronics, Massachusetts Institute of Technology, Cambridge, MA 02139, USA}
\def\LLaffil{MIT Lincoln Laboratory, Lexington, MA 02421, USA}
\def\Physaffil{Department of Physics, Massachusetts Institute of Technology, Cambridge, MA 02139, USA}
\def\EECSaffil{Department of Electrical Engineering and Computer Science, Massachusetts Institute of Technology, Cambridge, MA 02139, USA}
\def\IBMaffil{IBM Thomas J. Watson Research Center, Yorktown Heights, NY 10598, USA}
\def\affilAQ{\textit{Atlantic Quantum, Cambridge, MA 02139}}
\def\affilGoogle{\textit{Google Research, Mountain View, CA, USA.}}

\author{Leon Ding}
\email{leonding@mit.edu}
\affiliation{\Physaffil}
\affiliation{\RLEaffil}

\author{Max~Hays}
\affiliation{\RLEaffil}

\author{Youngkyu~Sung}
\altaffiliation[Present address: ]{\affilAQ}
\affiliation{\RLEaffil}
\affiliation{\EECSaffil}

\author{Bharath~Kannan}
\altaffiliation[Present address: ]{\affilAQ}
\affiliation{\RLEaffil}
\affiliation{\EECSaffil}

\author{Junyoung~An}
\affiliation{\RLEaffil}
\affiliation{\EECSaffil}

\author{Agustin~Di~Paolo}
\altaffiliation[Present address: ]{\affilGoogle}
\affiliation{\RLEaffil}

\author{Amir~H.~Karamlou}
\affiliation{\RLEaffil}

\author{Thomas~M.~Hazard}
\affiliation{\LLaffil} 

\author{Kate~Azar}
\affiliation{\LLaffil}

\author{David~K.~Kim}
\affiliation{\LLaffil}

\author{Bethany~M.~Niedzielski} 
\affiliation{\LLaffil} 

\author{Alexander~Melville} 
\affiliation{\LLaffil}

\author{Mollie~E.~Schwartz}
\affiliation{\LLaffil}

\author{Jonilyn~L.~Yoder}
\affiliation{\LLaffil}

\author{Terry~P.~Orlando}
\affiliation{\RLEaffil} 

\author{Simon~Gustavsson} 
\altaffiliation[Additional address: ]{\affilAQ}
\affiliation{\RLEaffil} 

\author{Jeffrey A. Grover}
\affiliation{\RLEaffil}

\author{Kyle Serniak} 
\affiliation{\RLEaffil}
\affiliation{\LLaffil} 

\author{William D. Oliver}
\email{william.oliver@mit.edu}
\affiliation{\Physaffil} 
\affiliation{\RLEaffil} 
\affiliation{\EECSaffil} 
\affiliation{\LLaffil} 

\date{\today}

\begin{abstract}
We propose and demonstrate an architecture for fluxonium-fluxonium two-qubit gates mediated by transmon couplers (FTF, for fluxonium-transmon-fluxonium).
Relative to architectures that exclusively rely on a direct coupling between fluxonium qubits, FTF enables stronger couplings for gates using non-computational states while simultaneously suppressing the static controlled-phase entangling rate ($ZZ$) down to kHz levels, all without requiring strict parameter matching. 
Here we implement FTF with a flux-tunable transmon coupler and demonstrate a microwave-activated controlled-Z (CZ) gate whose operation frequency can be tuned over a \SI{2}{GHz} range, adding frequency allocation freedom for FTF's in larger systems.
Across this range, state-of-the-art CZ gate fidelities were observed over many bias points and reproduced across the two devices characterized in this work.
After optimizing both the operation frequency and the gate duration, we achieved peak CZ fidelities in the 99.85-99.9\% range.
Finally, we implemented model-free reinforcement learning of the pulse parameters to boost the mean gate fidelity up to $99.922\pm0.009\%$, averaged over roughly an hour between scheduled training runs.
Beyond the microwave-activated CZ gate we present here, FTF can be applied to a variety of other fluxonium gate schemes to improve gate fidelities and passively reduce unwanted~$ZZ$ interactions.
\end{abstract}

\maketitle

\section{Introduction}
\topic{Talk about advances over the past decade with transmons}
Over the past two decades, superconducting qubits have emerged as a leading platform for extensible quantum computation. The engineering flexibility of superconducting circuits has spawned a variety of different qubits~\cite{Nakamura1999, Ioffe1999, Mooij1999, Koch2007, Manucharyan2009, Yan2016, Gyenis2021}, 
with an abundance of different two-qubit gate schemes~\cite{Majer2007, Rigetti2010, Poletto2012, Chow2013, Barends2014, Reagor2018, Campbell2020, Zhang2021, Sung2021}. 
More recently, tunable coupling elements have aided efforts to scale to larger multi-qubit systems~\cite{Yan2018}, including a demonstration of quantum advantage with 53 qubits~\cite{Arute2019} and quantum error correction improving with code distance in the surface code~\cite{Acharya2023}.
While there exists a large selection of superconducting qubits, almost all advancements toward processors at scale have been carried forward with the transmon qubit~\cite{Koch2007}, a simple circuit consisting of a Josephson junction in parallel with a shunt capacitance. However, that simplicity comes at a cost. A relatively large transition frequency~($\sim$\SI{5}{GHz}) makes the transmon more sensitive to dielectric loss~\cite{Smith2020}, and a weak anharmonicity~($\sim$\SI{-200}{MHz}) presents a challenge for both performing fast gates and designing multi-qubit processors.

\topic{Introduce fluxonium and talk about its advantages}
The fluxonium qubit~\cite{Manucharyan2009, Earnest2018, Nguyen2019} is a promising alternative to the transmon for gate-based quantum information processing~\cite{Nguyen2022}, which alleviates both of these drawbacks.
The fluxonium circuit consists of a capacitor (typically smaller than that of a transmon), a Josephson junction, and an inductor all in parallel. The transition frequency between the ground and first excited state of the fluxonium is usually between \SI{100}{MHz} and \SI{1}{GHz} at an external flux bias of~$\SI{0.5}{\Phi_0}$. At these low operating frequencies, energy relaxation times~$T_1$ exceeding~\SI{1}{ms}~\cite{Pop2014,Somoroff2021} have been observed, alongside anharmonicities of several~GHz.
With these advantages, fluxonium qubits have already achieved single-qubit gate fidelities above~99.99\%~\cite{Somoroff2021}.
Two-qubit gates relying on capacitive coupling~\cite{Bao2022, Dogan2022, Moskalenko2022, Ficheux2021, Xiong2022}, however, are more challenging because the same small transition matrix elements which improve~$T_1$ concomitantly reduce the interaction strength between qubits.
Moreover, direct capacitive coupling results in an undesired, always-on entangling rate~($ZZ$). 
In previous works, a variety of strategies were employed to reduce the~$ZZ$, such as keeping coupling strengths small or using ac-Stark drives, all of which have their own individual drawbacks. 
Finally, two-qubit gates must also reliably contend with frequency collisions with spectator qubits if they are to be scaled to larger arrays of qubits without sacrificing fidelity.
 
\topic{In this work, what do we do}
In this work, we introduce the fluxonium-transmon-fluxonium circuit (FTF) as a novel architecture for coupling fluxonium qubits via a transmon coupler~[\fref{fig:device}\panel{a}].
FTF suppresses the static~$ZZ$ down to kHz levels in a manner nearly insensitive to design parameter variations while simultaneously providing strong couplings for two-qubit gates via non-computational states.
Using FTF, we propose and demonstrate a microwave-activated CZ gate between two fluxonium qubits in a 2D-planar geometry.
This gate takes advantage of strong capacitive couplings which create a manifold of highly hybridized states that mix the first higher transition ($\ket{1} \leftrightarrow \ket{2}$) of both fluxonium qubits with the transmon's lowest transition ($\ket{0} \leftrightarrow \ket{1}$).
Despite these strong couplings, the computational states remain relatively unhybridized due to the large qubit-coupler detuning, allowing for high-quality single-qubit gates in addition to the two-qubit gate.
We applied microwave pulses to drive a full oscillation to and from this manifold contingent on both fluxonium qubits starting in their excited states and benchmarked an average CZ gate fidelity of up to 99.922~$\pm$ 0.009\% in \SI{50}{ns} via Clifford-interleaved randomized benchmarking.

The flux tunability of the transmon coupler also allows for the operation of the CZ gate at a wide range of microwave drive frequencies, providing a convenient way to avoid frequency collisions \textit{in situ} in larger-scale devices. 
Such \textit{in situ} tunability is critical for microwave-activated gates, as dependence on a particular frequency layout places an exponentially difficult demand on fabrication precision~\cite{Sheldon2016, Hertzberg2021} as the number of qubits increases. Our devices also exhibit up to millisecond~$T_1$ in a multi-qubit planar geometry, placing them among the highest coherence superconducting qubits to date and priming them for use in larger systems.

\section{FTF Architecture}

\begin{figure}[!htb]
\includegraphics{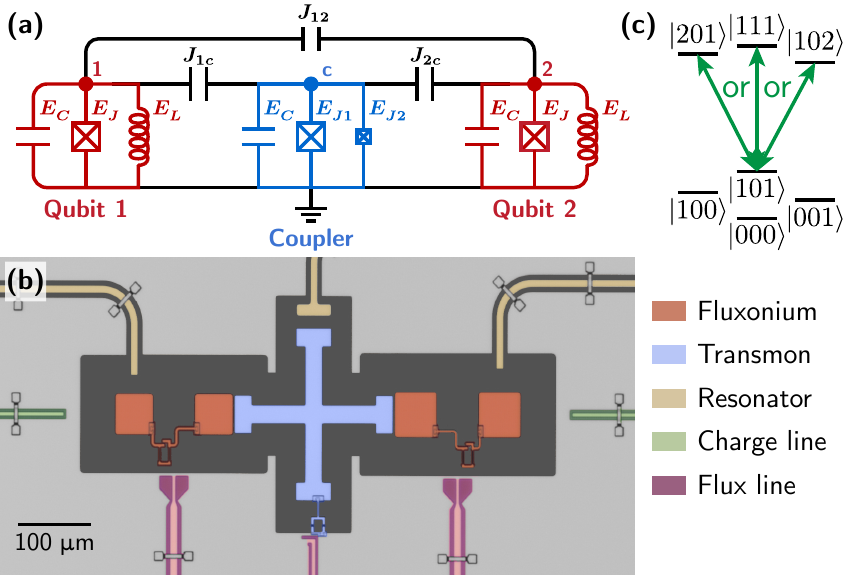}
\caption{\label{fig:device} \textbf{Device overview and gate principle.} \cpanel{a} Simplified circuit schematic of two fluxonium qubits (red) capacitively coupled to a tunable-transmon coupler (blue). \cpanel{b}  False-colored optical micrograph of the two fluxonium qubits  and the transmon along with their readout resonators, charge lines, and local flux lines. Arrays of 102 Josephson junctions in series form the fluxonium inductances. \cpanel{c} Energy level diagram illustrating the principle of the CZ gate. In practice, levels~$\ket{201}$,~$\ket{102}$, and~$\ket{111}$ are highly hybridized, and selectively driving any of these transitions results in a CZ gate.}
\end{figure}

\topic{Hamiltonian.}
Our device configuration consists of two differential fluxonium qubits capacitively coupled to a grounded transmon coupler, with a much weaker direct capacitive coupling between the two fluxonium qubits~[\fref{fig:device}\panel{b}]. The two lowest-lying states of each fluxonium form the computational basis~$\{\ket{00}, \ket{01}, \ket{10}, \ket{11}\}$, and the first excited state of the coupler, in addition to the second excited states of the fluxonium qubits, serve as useful non-computational states. Modeling only the qubits and their pairwise capacitive couplings, our system Hamiltonian is

\begin{eqnarray} \label{eq:hamiltonian}
\hat{H}/h &=& \sum_{i=1,2} 4 E_{C,i} \hat{n}^2_i + \frac{1}{2}E_{L,i}\hat\phi_i^2 - E_{J,i}\cos(\hat\phi_i - \phi_{\text{ext},i})\nonumber\\
 &+& 4 E_{C,c} \hat{n}_c^2 - E_{J1,c}\cos(\hat\phi_c) - E_{J2,c} \cos(\hat\phi_c - \phi_{\text{ext},c}) \nonumber\\
&+& J_{1c}\hat{n}_1 \hat{n}_c + J_{2c}\hat{n}_2\hat{n}_c + J_{12}\hat{n}_1\hat{n}_2,
\end{eqnarray}

\noindent where~$E_C$, $E_J$, and $E_L$ represent the charging, Josephson, and inductive energies, respectively.
Subscripts~${i=1, 2}$ index the two fluxonium nodes, and subscript~$c$ labels the coupler node.
Here we also introduced the external phase~$\phi_\text{ext}$, which is related to the external flux~$\Phi_\text{ext}$ through the expression~$\Phi_\text{ext}/\Phi_0 = \phi_\text{ext} / 2\pi$ for each qubit.
For both our main device~(Device A) and a secondary device (Device B), the experimentally obtained Hamiltonian parameters are listed in \tref{tab:hamiltonian}, along with the measured coherence times.

\begin{table*}[!htb]
\caption{\label{tab:hamiltonian}
\textbf{Characterization of FTF devices.} Hamiltonian parameters for both Device \textbf{A} and Device \textbf{B} were obtained by fitting two-tone spectroscopy data and the static~$ZZ$ rate vs. coupler flux.
Coherence times were measured by biasing each fluxonium at~$\Phi_\text{ext} = \SI{0.5}{\Phi_0}$ using only the global flux bias.
Unless otherwise stated, all data in this manuscript corresponds to Device A.
}
\begin{ruledtabular}
\begin{tabular}{lcccccccccc}
& &
$E_C$ (GHz)&
$E_L$ (GHz)&
$E_J$ (GHz)&
$N_{JJ}$ &
$\omega_{01}/2\pi$ (GHz) &
$\omega_r/2\pi$ (GHz)&
$T_1$ (\SI{}{\micro s})&
$T_2^R$ (\SI{}{\micro s})&
$T_2^E$ (\SI{}{\micro s}) \\
\colrule
\textbf{A} &Fluxonium 1 & 1.41 & 0.80 & 6.27 & 102 & 0.333 & 7.19 & 560 & 160 & 200\\
\textbf{A} &Fluxonium 2 & 1.30 & 0.59 & 5.71 & 102 & 0.242 & 7.08 & 1090 & 70 & 190\\
\textbf{A} &Transmon c & 0.32 &  & 3.4, 13 & -- & -- & 7.30 & -- & -- & --\\
\textbf{B} &Fluxonium 1 & 1.41 & 0.88 & 5.7 & 102 & 0.426 & 7.20 & 450 & 230 & 240\\
\textbf{B} &Fluxonium 2 & 1.33 & 0.60 & 5.4 & 102 & 0.281 & 7.09 & 1200 & 135 & 310\\
\textbf{B} &Transmon c & 0.30 &  & 3.0, 13 & -- & -- & 7.31 & -- & -- & --\\
\hline
\hline
& &\multicolumn{3}{c}{$J_{1c}$ (MHz)} & \multicolumn{3}{c}{$J_{2c}$ (MHz)}  & \multicolumn{3}{c}{$J_{12}$ (MHz)} \\
\colrule
\textbf{A} &Coupling strengths & \multicolumn{3}{c}{570} & \multicolumn{3}{c}{560}  & \multicolumn{3}{c}{125} \\
\textbf{B} &Coupling strengths & \multicolumn{3}{c}{550} & \multicolumn{3}{c}{550}  & \multicolumn{3}{c}{120} 
\end{tabular}
\end{ruledtabular}
\end{table*}

\subsection{Gate principles}

\topic{Explain the coupling and higher levels for gate. Relevant detunings to sideband transitions}
The operating principles of FTF are fundamentally different than those of all-transmon circuits~\cite{Yan2018}.
Due to its relatively high frequency, the coupler interacts negligibly with the computational states of the qubits.
Instead, the coupler predominantly interacts with the higher levels of the qubits, acting as a resource for two-qubit gates without adversely affecting single-qubit gates. 

We describe the quantum state of the system using the notation~$\ket{jkl}$, where~$j$, $k$, and~$l$ denote the energy eigenstates in the uncoupled basis of fluxonium~1, the coupler, and fluxonium~2, respectively.
While coupler-based gates are often activated by baseband flux pulses, here we generate the required entangling interaction via a microwave pulse from~$\ket{101}$ to a non-computational state of the joint system. 
As illustrated in \fref{fig:device}\panel{c}, a single-period Rabi oscillation from~$\ket{101}$ to either~$\ket{201}$, $\ket{111}$, or~$\ket{102}$ gives the~$180^\circ$ conditional phase shift necessary for a CZ gate, provided no other transitions are being driven. 
We note that, throughout this work, we label the eigenstates according to their maximum overlap with the uncoupled qubit/coupler states at~$\Phi_\text{ext,c} = 0$ (equivalently ${\Phi_\text{ext,c} = \SI{1}{\Phi_0}}$).
We perform this labeling at~${\Phi_\text{ext,c} = 0}$ because tuning the coupler flux results in avoided crossings among the higher levels of the system. 

\topic{Advantages of FTF}
In general, stronger coupling strengths result in larger detunings from parasitic transitions in two-qubit gate schemes, yet doing so often results in unintended consequences. Two common drawbacks of larger coupling strengths are remedied using the FTF architecture: (1)~crosstalk due to non-nearest-neighbor couplings, and (2)~unwanted static~$ZZ$ interactions. In all transmon-based architectures, the same level repulsion that enables the two-qubit gate also creates level repulsions within the computational subspace. This is because all transition frequencies and charge matrix elements of adjacent levels in a transmon have similar values. As this hybridization among the computational states increases, charge drives will produce non-local microwave crosstalk to unwanted qubit transitions. In FTF, the large ratio of transition matrix elements~$|\bra{2}\hat{n}\ket{1}| / |\bra{1} \hat{n} \ket{0}|$ for fluxonium qubits, the large fluxonium anharmonicity, and the large detunings between the transmon and each fluxonium all serve to mitigate these negative side-effects. 

\subsection{\textit{ZZ} reduction}

\noindent \topic{ZZ Introduction}
Formally, the~$ZZ$ interaction rate is defined as~${\zeta = (E_{11} - E_{10} - E_{01} + E_{00})/h}$ (for two qubits) or~${\zeta = (E_{101} - E_{100} - E_{001} + E_{000})/h}$ (for two qubits and a coupler). It describes an unwanted, constant, controlled-phase-type entangling rate caused by the collective level repulsions from the many non-computational states of superconducting qubits acting on the computational states. 

\begin{figure}[!htb]
\includegraphics{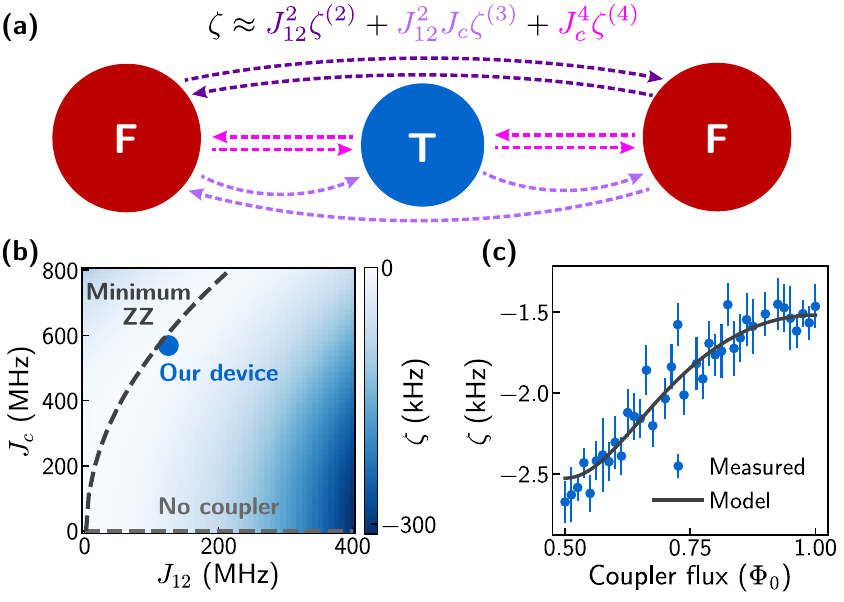}
\caption{\label{fig:zz} \textbf{\textit{ZZ}-reduction in the FTF architecture.} \cpanel{a}~A perturbative treatment of the couplings~$J_{ij}$ shows energy shifts in FTF to be dominated by virtual transitions (dashed arrows) of second (dark purple), third (light purple), and fourth (pink) order between the fluxonium qubits (maroon circles) and the coupler (blue circle). \cpanel{b}~Numerical simulation of~$\zeta$ as a function of~$J_c = J_{1c} = J_{2c}$ and~$J_{12}$ with the experimentally extracted qubit parameters. With the coupler, a ratio of coupling strengths always exists that minimizes~$\zeta$ (dark gray). \cpanel{c}~Measured and simulated~$\zeta$ as a function of the coupler flux for the experimental device parameters. The~$ZZ$ rate remains nearly constant between~$-1.5$ and \SI{-2.7}{kHz}.}
\end{figure}

\topic{Perturbative analysis of ZZ}
A key feature of the FTF architecture is its ability to suppress $\zeta$, despite the strong coupling strengths that would typically amplify it. This low~$\zeta$ can be understood by considering the couplings~$J_{ij}$ perturbatively up to fourth order.
At each order~$m$, the perturbative correction can be considered an~$m$th-order virtual transition between the states of the uncoupled qubits; the strength of a particular transition is proportional to the product of the corresponding couplings~$J_{ij}$.
In \fref{fig:zz}\panel{a}, we illustrate the dominant virtual transitions up to fourth order: the first-order correction is zero; at second order, only direct transitions between the two fluxonium qubits contribute to~$\zeta$; at third order, the only allowed transitions form three-cycles between the three qubits; and at fourth order, we find that transmon-mediated transitions between the two fluxonium qubits dominantly contribute to~$\zeta$. As such, we can write~$\zeta$ to fourth order as 

\begin{equation}\label{eq:zz_pert}
\zeta \approx J_{12}^2\zeta^{(2)} + J_{12}J_c^2\zeta^{(3)} + J_c^4\zeta^{(4)},
\end{equation}

\noindent where~$\zeta^{(i)}$ depend only on the uncoupled states, and we assume~$J_{1c} = J_{2c} = J_c$. Specifically, we find our device to be well-described by~$\zeta^{(2)} = \SI{-2.1e-3}{GHz^{-1}}$, $\zeta^{(3)} =  \SI{1.4e-3}{GHz^{-2}}$, and~$\zeta^{(4)} = \SI{-2.6e-4}{GHz^{-3}}$ at~$\Phi_\text{ext,c} = \SI{0.5}{\Phi_0}$. Critically, both the second- and fourth-order terms are negative, while the third-order term is positive. This is a direct consequence of the perturbation theory: for virtual transitions to excited states above the computational subspace, even-order terms describe level repulsions, while odd-order terms describe level attractions. The relatively low~$ZZ$ in the FTF system stems from this cancellation between even and odd terms (see \aref{appendix:ZZ} for further details on the $ZZ$ cancellation).

\topic{Minimum ZZ trajectory and value}
To understand this quantitatively, we numerically calculate~$\zeta$ by diagonalizing \eref{eq:hamiltonian} as a function of~$J_c$ and~$J_{12}$ [\fref{fig:zz}\panel{b}].
We find that~$\zeta$ can be almost perfectly canceled by appropriate choices of the couplings, as traced by the darker dashed line. Within perturbation theory, this curve of minimum~$\zeta$ is a parabola:~$d\zeta /d J_{12} = 0 \rightarrow J_{12} = -J_{c}^2 \zeta^{(3)}/2\zeta^{(2)}$. By inserting this expression into \eref{eq:zz_pert}, we obtain the dependence of~$\zeta$ \textit{along} the parabola~$\zeta_\mathrm{min} = J_c^4(- \zeta^{(3)}\zeta^{(3)}/4\zeta^{(2)} + \zeta^{(4)})$. For our device parameters, the two terms in parentheses almost cancel, summing to~$-2 \times 10^{-6}~\mathrm{GHz}^{-3}$.

\topic{ZZ vs coupling strengths}
Importantly, $|\zeta|$ remains below \SI{10}{kHz} for~$J_c$ values of up to~\SI{1}{GHz}, while maintaining the optimal coupling ratio. 
To take full advantage of this phenomenon, we designed the coupling strengths to be as large as reasonable for our geometry.
Furthermore, by choosing the transmon to have a grounded geometry, a near-optimal ratio of~$J_{c}^2 / J_{12}$ is maintained while freely varying the fluxonium-transmon capacitance (see \aref{appendix:grounded_vs_floating} for full details).
In addition to this insensitivity to the underlying coupling capacitance, $\zeta$ is only weakly dependent on~$J_c$, and~$J_{12}$: independent errors in~$J_c$ and~$J_{12}$ by up to 20\% would increase~$\zeta$ in our device by a maximum of \SI{11}{kHz} (modeling the worst case scenario in which~$J_{12}$ increases and~$J_c$ decreases).
Such robustness will be critical in larger-scale devices, as capacitive coupling strengths cannot be changed after device fabrication and are subject to fabrication variations.

\topic{ZZ vs coupler frequency}
The value of~$\zeta$ is also insensitive to the coupler frequency, allowing us to safely bias the system at any~$\Phi_\text{ext,c}$. This is unsurprising, as the coupler energy levels are far from any resonances with the computational states. In other words, any change in the coupler frequency must compete with the large detuning between the coupler and fluxonium~$\ket{0}\leftrightarrow \ket{1}$ transitions. To validate our models, we experimentally determined~$\zeta$ by measuring the frequency of fluxonium~1 using a Ramsey experiment while preparing fluxonium~2 in the ground or excited state. Taking the difference in fitted frequencies associated with the two initial state preparations yields the experimental value of~$\zeta$, which we find to closely follow our numerical simulations as a function of the coupler flux [see \fref{fig:zz}\panel{c}]. 
An alternative approach to~$ZZ$ reduction with fluxonium qubits is to apply always-on ac-Stark drives~\cite{Xiong2022, Dogan2022}. While this is an effective means to reduce~$ZZ$ in few-qubit devices, the requirement of an additional calibrated drive per qubit becomes increasingly prohibitive as system sizes grow. 

\section{Gate Calibration}

\topic{Readout}
Before discussing our single- and two-qubit gates, we first describe the qubit readout and flux biasing. 
In thermal equilibrium, our fluxonium qubits have nearly equal populations in the ground and excited states~${(k_\mathrm{B}T > \hbar \omega_{01})}$. To address this, we initialized each qubit in either the ground or excited state at the beginning of each experiment via projective readout and heralded the desired initial state~\cite{Johnson2012} (see \aref{appendix:readout} for further details). To realize independent qubit initialization in our system, each qubit was capacitively coupled to a separate readout resonator, allowing us to perform high-fidelity, single-shot readout within the full computational basis. All three resonators were further coupled to a common Purcell filter~\cite{Sete2015}. 

\topic{Specific wiring and layout}
We used a global biasing coil to tune the flux across the entire device and used additional local flux lines biased through coaxial cables for independent control of each qubit. This allowed us to freely change the coupler flux while holding each fluxonium at~$\Phi_\text{ext} = 0.5 \Phi_0$. Although only DC flux was required in our experiment, our device is fully compatible with fast-flux pulses. As such, FTF presents an opportunity  to investigate iSWAP, Landau-Zener, or other flux-modulated gates in a system with low static~$ZZ$ rates~\cite{Bao2022, Campbell2020, Zhang2021, Setiawan2021, Setiawan2022, Mundada2020}.  

\topic{Talk about flux lines}
Unfortunately, we found that our qubit coherence times were sensitive to bias-induced heating from the coaxial flux lines. We have found that both qubit~$T_1$ and~$T_2$ (both Ramsey and spin-echo) drop with increasing current. To minimize this effect when performing two-qubit experiments, the global coil was used to simultaneously bias the two fluxonium qubits as close as possible to their operation points~${\Phi_\text{ext,1} = \Phi_\text{ext,2} = \SI{0.5}{\Phi_0}}$. The local flux lines were then used to more precisely tune~$\Phi_\text{ext,1}$ and~$\Phi_\text{ext,2}$ and bias the coupler flux. We emphasize that this nonideality is not fundamental nor unique to the FTF architecture and can be improved in future experiments by optimized construction and filtering of our flux-bias lines.

\subsection{Single-qubit gates}

\topic{General setup}
To deconvolve the aforementioned heating effects from the measurement results, we biased the qubits solely with the global coil when characterizing individual qubit coherences (see \tref{tab:hamiltonian}). Notably, fluxonium~2 in our device achieves a lifetime of over a millisecond, with similar performance reproduced in Device B (see \aref{appendix:other_device}). In accordance with the higher qubit frequency, fluxonium~1 has a shorter lifetime, and the~$T_2^E$ of all characterized qubits peaks between 200-\SI{300}{\micro s}, likely limited by photon-shot noise from occupation of the resonator or filtering of the flux lines.

\begin{figure}[!htb]
\includegraphics{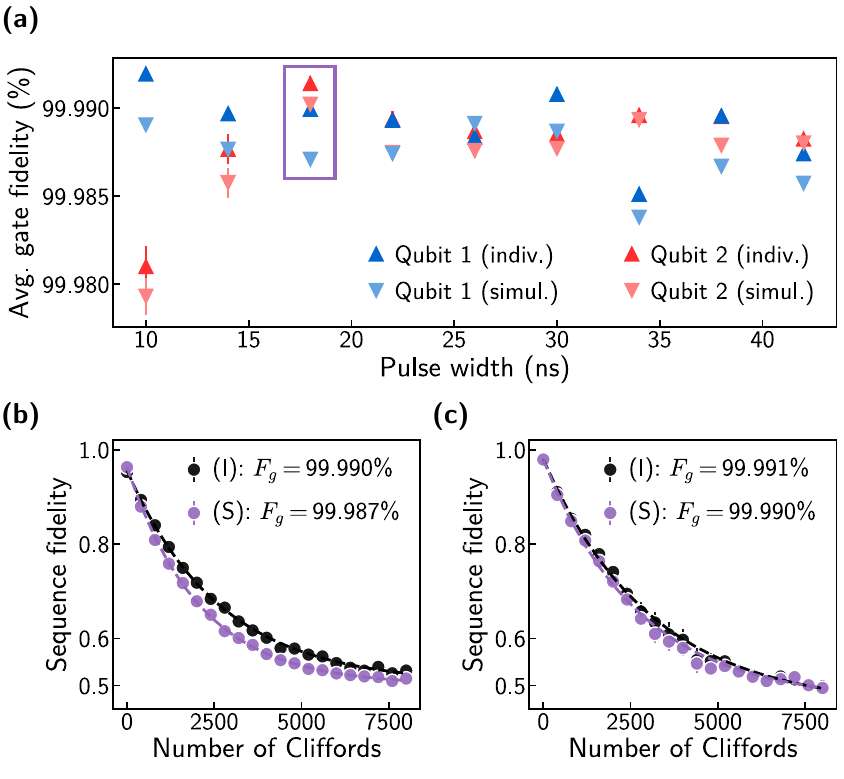}
\caption{\label{fig:1qb} \textbf{Single-qubit benchmarking on our multi-qubit device.} \cpanel{a} Average single-qubit gate fidelities obtained by individual and simultaneous Clifford randomized benchmarking as a function of the pulse width. Error bars for the majority of data points are within the size of the marker and correspond to standard errors about the mean. \cpanel{b-c} Individual (I) and simultaneous (S) RB traces of an \SI{18}{ns} gate (purple box in (a)) for qubit 1 and qubit 2, respectively. Individual and simultaneous average gate fidelities have a standard error of about $3\times 10^{-6}$ for qubit 1 and $4\times 10^{-6}$ for qubit 2. The larger uncertainties in the qubit 2 data arise from coherent errors, which begin to dominate for gates shorter than \SI{18}{ns} (red points in \panel{a}).}
\end{figure}

\topic{Single-qubit gate results}
We realized single-qubit gates by calibrating Rabi oscillations generated by a resonant charge drive using a cosine pulse envelope.
To quantify the fidelities of these gates, we performed individual as well as simultaneous Clifford randomized benchmarking (RB) using a microwave-only gate set, 
${\{I,\, \pm X,\, \pm Y,\, \pm X_{\pi/2},\, \pm Y_{\pi/2}\}}$, to generate the Clifford group, resulting in an average of 1.875 gates per Clifford~\cite{Magesan2011, Barends2014}.
In our decomposition, all gates had an equal time duration and were derived from a single calibrated~$X_\pi$ pulse by halving the amplitude and/or shifting its phase (see \aref{appendix:single_qubit} for the full calibration sequence).
Here, both qubits were biased at $\Phi_\text{ext} = \SI{0.5}{\Phi_0}$, with minimal current through the coupler flux line.
In \fref{fig:1qb}\panel{a}, we varied the pulse width from \SI{10}{ns} to \SI{42}{ns} and found average single-qubit gate fidelities consistently near or above 99.99\% and show the explicit randomized benchmarking traces in \fsref{fig:1qb}\panel{b-c} for a gate duration of \SI{18}{ns}. In this range, the incoherent error begins to trade off with the coherent error (from violating the rotating wave approximation), with qubit~1 able to tolerate faster gates than qubit~2 due to its higher frequency. We note gates were additionally calibrated at \SI{6}{ns} with significantly lower fidelities on both qubits due to large coherent errors. Overall, our fidelities from simultaneously applied gates are $<$5$\times$10$^{-5}$ lower than the individually applied ones, a testament to the low~$ZZ$ rate measured in our system. We suspect the small difference in fidelity is caused by microwave crosstalk between charge lines and qubits.

\subsection{Two-qubit CZ gate}
\begin{figure*}[!htb]
\includegraphics{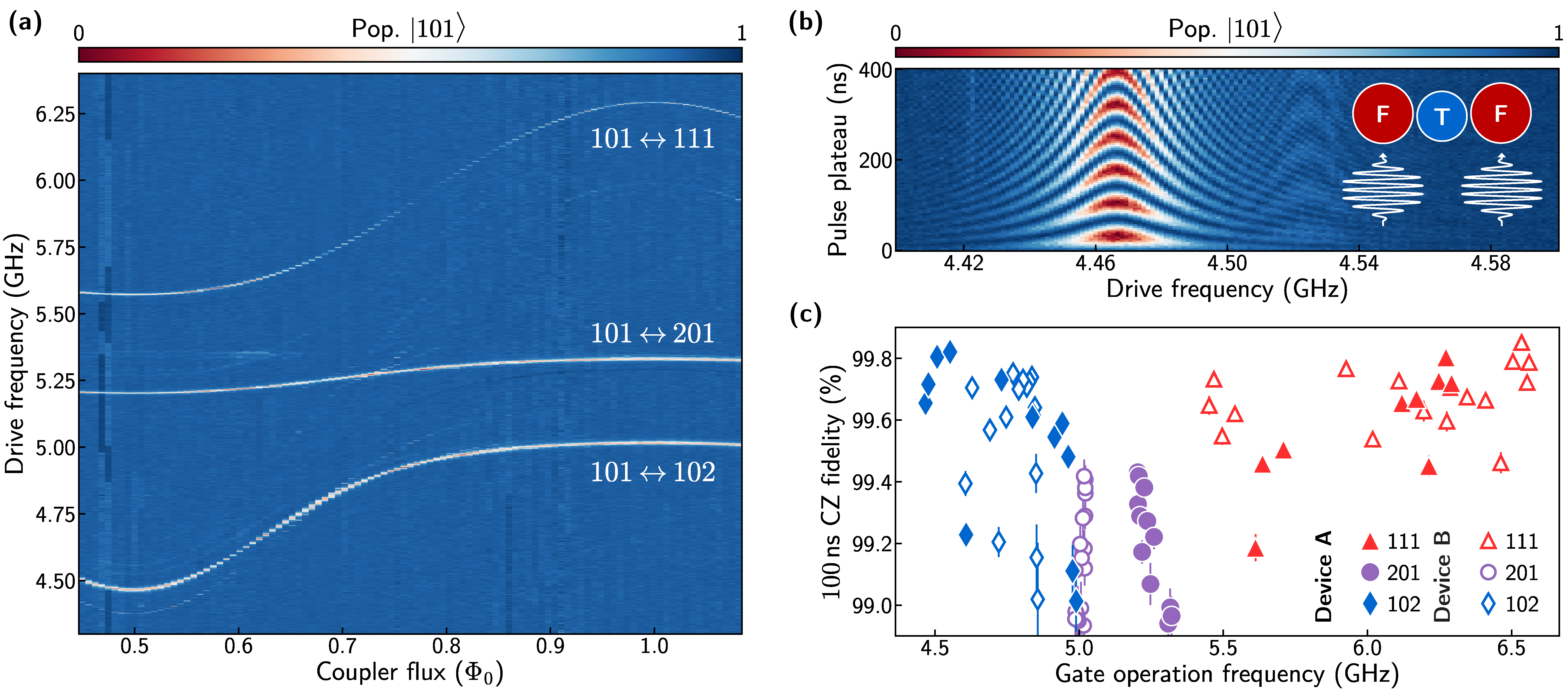}
\caption{\label{fig:map_spec} \textbf{CZ gate as a function of flux.} \cpanel{a} Spectroscopy of relevant non-computational states. The qubits were initialized in the~$\ket{101}$ state via single-shot readout. \cpanel{b} Time-domain Rabi oscillations of the~$\ket{101} \leftrightarrow \ket{102}$ transition as a function of the drive frequency. The faint chevron pattern at \SI{4.52}{GHz} arises from the~$\ket{001}\leftrightarrow \ket{002}$ transition and was visible due to imperfect state initialization. In all plots, populations have been normalized to between 0 and 1. Inset shows that gates were driven with simultaneous phase-locked pulses applied to each fluxonium charge line. \cpanel{c} CZ gate fidelities using a fixed~\SI{100}{ns} cosine pulse envelope, driving each transition in \panel{a} across the entire~${0.5-\SI{1}{\Phi_0}}$ range, linearly sampled over 21 points. A secondary device (Device B) with slightly different Hamiltonian parameters supports the reliability of our architecture. Points with fidelity below 98.9\% correspond to failures in the automated calibration and are therefore excluded from the plot. All gate fidelities were obtained from interleaved randomized benchmarking averaged over different 20 randomizations, with error bars corresponding to the standard error.}
\end{figure*}

\topic{Spectroscopy and transition selection}
We began our investigation of the two-qubit CZ gate by performing spectroscopy of the relevant non-computational state transitions.
With the system initialized in~$\ket{101}$, we swept the coupler flux~$\Phi_\text{ext,c}$ to map out the transition frequencies to the three dressed states~$\ket{201}$, $\ket{111}$, and~$\ket{102}$ [see \fref{fig:map_spec}\panel{a}].
Importantly, we found that~$\ket{111}$ crosses both~$\ket{201}$ and~$\ket{102}$ (at~$\Phi_\text{ext,c} \approx \SI{0.65}{\Phi_0}$), with an avoided crossing strength of nearly \SI{1}{GHz}.
With such strong hybridization, a high-performance gate could be realized by driving any of the three energy levels over a wide coupler flux range. Nevertheless, the transitions yielded varying performance depending on their coherence times and the proximity of undesired transitions, whose frequencies we extracted in the same measurement by heralding different initial states (see \aref{appendix:transitions}).

\topic{Gate driving}
We activated the gate interaction associated with each transition by simultaneously applying a charge drive to each fluxonium [\fref{fig:map_spec}\panel{b} inset].
These drives were chosen to have equal amplitude, with a relative phase between them to maximize constructive interference at the intended transition.
We found that using two constructive drives was a convenient method for reducing the total applied power for a given Rabi rate, resulting in a reduced ac-Stark shift from off-resonant transitions. 
In severe cases, a large ac-Stark shift could prevent the realization of a 180$^\circ$ conditional phase and increase leakage into non-computational states.
An alternative approach exists to tune the relative phase and amplitude of the two drives to result in complete destructive interference of the nearest leakage transition (see \aref{appendix:microwave_crosstalk}).
While theoretically we expect the relative phase and amplitude to be important for reducing leakage, we found that as long as a~$180^\circ$ conditional phase was attainable, the CZ gate fidelity was relatively insensitive to these two parameters.
\figureref{fig:map_spec}\panel{b} shows the familiar Rabi chevrons when the transition was driven as a function of frequency; in experimental practice, our two-qubit gate is quite similar to driving single-qubit Rabi oscillations.

\topic{Gate calibration and characterization}
For a given transition and pulse duration, there are four critical parameters associated with the CZ gate to calibrate: (1) the overall drive amplitude of the two phase-locked drives to ensure a single-period oscillation, (2) the drive frequency to ensure a~$180^\circ$ conditional phase accumulation, and (3-4) the single-qubit phases accumulated on each fluxonium during the gate.
These single-qubit phase corrections can be conveniently implemented through virtual-Z gates by adjusting the phases of subsequent single-qubit gates~\cite{Mckay2017}.
After calibration, we extract the gate fidelity by performing Clifford interleaved randomized benchmarking, averaging over 20 different randomizations~\cite{Corcoles2013, Magesan2012, Barends2014}.
Similar to our single-qubit Clifford decomposition, we generated the two-qubit Clifford group with the gate set 
${\{I,\, \pm X,\, \pm Y,\, \pm X_{\pi/2},\, \pm Y_{\pi/2},\, \text{CZ}\}}$, yielding an average of 8.25 single-qubit gates and 1.5 CZ gates per Clifford.
A more detailed gate calibration and characterization description can be found in \aref{appendix:two_qubit}.

\topic{Gate vs. Flux}
The FTF approach offers a potential solution to frequency-crowding by allowing for an adjustable gate operation frequency.
To demonstrate this frequency-flexibility in our device, we linearly sampled the coupler flux~$\Phi_\text{ext,c}$ at 21 values between~$\SI{0.5}{\Phi_0}$ and~$\SI{1}{\Phi_0}$ and calibrated a CZ gate across all three transitions in \fref{fig:map_spec}\panel{a} while maintaining a constant \SI{100}{ns} pulse length [\fref{fig:map_spec}\panel{c}].
Each data point in \fref{fig:map_spec}\panel{c} represents a fully automated re-calibration of all single- and two-qubit gate parameters without manual fine-tuning.
Missing points indicate either failed calibrations or fidelities lower than~$98.9\%$, which may be caused by TLS or nearly resonant unwanted transitions (see \aref{appendix:rb_vs_flux} for further details).
While these fidelities remain unoptimized over the pulse width, they indicate the ease and robustness of the tuneup, as well as the accessibility of state-of-the-art gate fidelities at a variety of drive frequencies.
We further emphasize this by including fidelities from Device B, a second fully characterized device with similar performance. While designed to be nominally identical, the non-computational states in Device B differ by up to \SI{300}{MHz} from Device A, with no significant detriment to the gate fidelities (see \aref{appendix:other_device} for additional characterization of Device B).

\begin{figure}[!htb]
\includegraphics{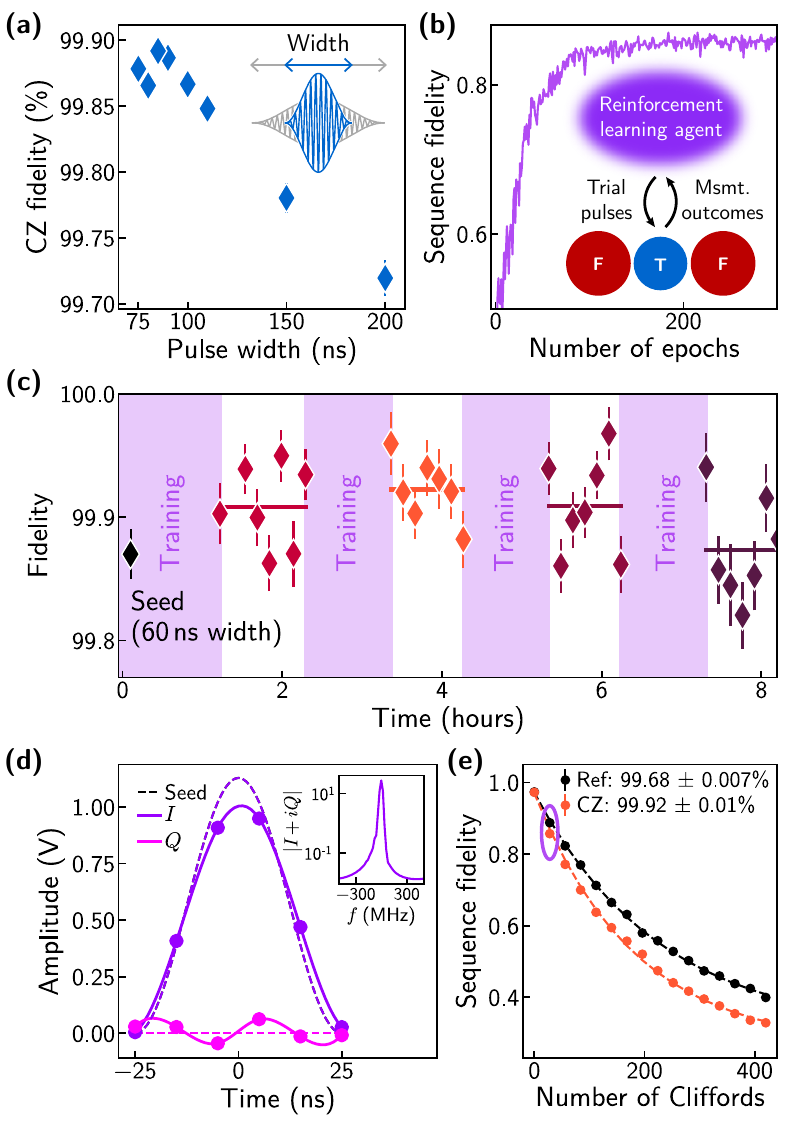}
\caption{\label{fig:fidelities} \textbf{Optimizing the CZ gate fidelity via reinforcement learning.} \cpanel{a} Gate fidelities as a function of the width of the cosine pulse envelope, averaged over 20 randomizations. The~$\ket{101}\leftrightarrow \ket{102}$ transition at~$\Phi_\text{ext,c} \approx \SI{0.575}{\Phi_0}$ was used for these gates. \cpanel{b} Fidelity of an interleaved randomized benchmarking sequence with 28 Cliffords using trial CZ gates sampled from the policy of a model-free reinforcement learning agent. After each epoch, the measurement results were used to update the agent's policy according to the PPO algorithm. 
\cpanel{c} In a full training run, the agent was first seeded with a \SI{50}{ns} cosine pulse, with an amplitude determined by a physics-calibrated \SI{60}{ns} gate (black diamond) but scaled by 60/50.  Then, the agent was trained to optimize the sequence fidelity of the \SI{50}{ns} pulse (b). The learned gate was then repeatedly evaluated using interleaved randomized benchmarking averaged over 10 randomizations. The next round of training was seeded with the optimized pulse from the previous round. Horizontal bars indicate the averaged fidelity after each round of training. 
\cpanel{d} Optimized pulse shape as learned by the agent. The agent was given control over six evenly-spaced $I(t)$ and $Q(t)$ voltage points (colored circles), with the pulse in between points determined by a cubic interpolation. Inset shows the Fourier transform of $I(t) + iQ(t)$. 
\cpanel{e} Reference and interleaved randomized benchmarking curves averaged over all 70 randomizations after the second round of training (orange points in \panel{c}). The purple circle at 28 Cliffords indicates the sequence length used for learning. All uncertainties in this figure correspond to the standard error of the mean.
}
\end{figure}

\topic{RB vs pulse width}
To investigate the trade-off between coherent and incoherent error, we characterized the gate fidelity as a function of the pulse width.
In general, all fidelities in \fref{fig:map_spec}\panel{c} may be improved by optimizing over this pulse width, with our highest fidelity gates using the~${\ket{101}\leftrightarrow \ket{102}}$ transition at~$\Phi_\text{ext,c} = 0.575 \Phi_0$ [\fref{fig:fidelities}\panel{a}].
At the observed optimal gate time of \SI{85}{ns} for this transition, we benchmarked a CZ fidelity of~${99.89\%\pm0.02\%}$. 
For longer gate durations, the gate error is dominated by the lifetime of the driven non-computational state, measured to be around \SI{10}{\micro s} at this transition and, in general, varied from~$5-\SI{20}{\micro s}$ across all transitions in \fref{fig:map_spec}\panel{a}.
The coherence times of the computational states also reduce the fidelity, but for our devices this error is negligible compared to the~$T_1$ of the non-computational state (see \aref{appendix:error} for an analytic error model of this gate process). 
At shorter gate lengths, coherent leakage into non-computational levels dominates the error, but due to the extreme degree of hybridization of the non-computational states and their subpar readout, we were not able to experimentally measure the location of the leaked population.

\topic{Reinforcement Learning}
To further improve the gate fidelity, we deployed a model-free reinforcement learning agent~\cite{Schulman2017,an2019deep,niu2019universal,Wang2020}, closely following the protocol described by Sivak \textit{et al.}~\cite{Sivak2022}.  
While reinforcement learning could not mitigate the incoherent errors dominating the gate at longer pulse widths, at shorter gate times ($<$\SI{70}{ns}) we found that it did offer an improvement via fine adjustments of the pulse parameters. 
To train the agent, we first seeded it with a physics-based pulse calibration with a pulse width of \SI{60}{ns}; our physics-based calibration failed at gate times less than this. 
Then, with a fixed CZ pulse width of \SI{50}{ns}, the agent was trained to maximize the sequence fidelity of interleaved randomized benchmarking at 28 Cliffords with a fixed random seed by optimizing the pulse shape and virtual-Z gates~[\fref{fig:fidelities}\panel{b}]. 
After each round of training, the optimized pulse was repeatedly evaluated by 
performing interleaved randomized benchmarking over 70 total Clifford sequences~[\fref{fig:fidelities}\panel{c}]. 
Training was then repeated using the optimized pulse shape from the previous training round as the seed for the next. 
For the training run shown in~\fref{fig:fidelities}\panel{c}, the fidelity peaked after the second round of training (orange points), with a time-averaged value of $99.922\pm0.009\%$.  
A Wilcoxon signed-rank test gives 97\% confidence that this mean is above 99.90\%.
As the run progressed, the average fidelity was observed to degrade. 
We hypothesize that this was due to system drifts beyond what the agent was able to mitigate. 

In the data shown in \fsref{fig:fidelities}\panel{b-e}, the agent was given full control of the $I$ and $Q$ quadratures of the pulse envelope as well as the single-qubit virtual-Z rotation angles. 
$I(t)$ and $Q(t)$ were discretized into six points equally spaced in time [\fref{fig:fidelities}\panel{d}] with a cubic interpolation determining the remaining points. 
Perhaps the most distinctive feature of the learned pulse is the shape of $Q(t)$: while the agent was seeded with $Q(t) = 0$, the learned pulse shape displays a distinct oscillation in $Q(t)$. 
Although we do not fully understand the origin of this oscillation, some intuition may be gained by examining the Fourier transform of the pulse shape [\fref{fig:fidelities}\panel{d} inset]. 
In the frequency domain, the oscillation in $Q(t)$ results in a distinct asymmetry of the pulse shape: at positive detunings from the carrier frequency, the spectral weight is suppressed, and vice versa for negative detunings. 
We hypothesize that this learned pulse asymmetry mitigates the effects of the nearest undesired transition, which is detuned by $\sim$+\SI{65}{MHz} at this bias point. 
However, we also note that attempts to mitigate the effects of this undesired transition using more established pulse shaping techniques did not improve the gate fidelity~\cite{Motzoi2009}.

\section{Outlook}

\topic{Summary of good thing on FTF}
Our work demonstrates an architecture in which high-fidelity, robustness against parameter variations, and extensibility are simultaneously realized.
We observed millisecond fluxonium lifetimes despite couplings to neighboring qubits, resonators, flux lines, and charge lines, all within a 2D-planar architecture.
Both the single- and two-qubit gates performed here are also simple -- operating on the basis of a Rabi oscillation.
The relative simplicity of this two-qubit gate was afforded by the FTF Hamiltonian and yielded a high fidelity operation across a large frequency-tunable range, reproduced across multiple devices.

\topic{ZZ, and benefits for over gate schemes}
One of the most notable features of the FTF scheme is the capacity for large coupling strengths while simultaneously reducing the~$ZZ$ interaction strength to kHz levels.
This is all done without strict parameter matching or additional drives. In fact, even computational state gates such as iSWAP or cross-resonance can benefit from FTF by utilizing this~$ZZ$ reduction without worrying about additional complications for single- and two-qubit gates. A fixed-frequency transmon (or simply a resonator) would suffice for this use case (see \aref{appendix:ZZ}).

\topic{Future improvements}
Despite already high gate fidelities, many avenues exist for improvement. First and foremost, the device heating when DC biasing qubits to their simultaneous sweet spot and tuning the coupler flux reduces the coherence times of our qubits.
By optimizing the mutual inductance between the flux lines and the qubits and improving the thermalization and filtering of the flux lines, we anticipate improvements in future experiments.
Even in the absence of local heating, we estimate a photon-shot-noise limit of~${T_2\sim \SI{400}{\micro s}}$, assuming an effective resonator temperature of~$T_\text{eff} = \SI{55}{mK}$~\cite{YanCampbell2018}.
Simply decreasing~$\chi$ and~$\kappa$ should increase this~$T_2$ limit at the expense of readout speed, which could be a worthwhile exchange in the high~$T_1$, low~$T_2$ limit. 

As is typical of fluxonium gates involving the non-computational states, the largest contribution to gate infidelity was the coherence of the~$\ket{201}$, $\ket{111}$,~$\ket{102}$ manifold. Yet, the lifetimes of these states were much lower than expected, given coherence times measured on transmons with similar frequencies. By optimizing regions of high electric field density that exist in our current design (notably the small fluxonium-transmon capacitor gap), we expect to improve these coherence times as well.

\topic{Outlook for fluxonium quantum computing}
While fluxonium has long exhibited impressive individual qubit performance, our work demonstrates a viable path forward for fluxonium-based large-scale processors capable of pushing the boundaries of noisy intermediate-scale quantum computing. 

\begin{acknowledgments}

L.D. is grateful to Devin Underwood for his continued support throughout this project, especially during the lockdown years. M.H. acknowledges useful discussions with Vlad Sivak. This research was funded in part by the U.S. Army Research Office Grant W911NF-18-1-0411, and by the Under Secretary of Defense for Research and Engineering under Air Force Contract No. FA8702-15-D-0001. L.D. gratefully acknowledges support from the IBM PhD Fellowship, Y.S., and J.A. gratefully acknowledge support from the Korea Foundation for Advances Studies, and B.K. gratefully acknowledges support from the National Defense Science and Engineering Graduate Fellowship program. M.H. was supported by an appointment to the Intelligence Community Postdoctoral Research Fellowship Program at the Massachusetts Institute of Technology administered by Oak Ridge Institute for Science and Education (ORISE) through an interagency agreement between the U.S. Department of Energy and the Office of the Director of National Intelligence (ODNI). Any opinions, findings, conclusions or recommendations expressed in this material are those of the author(s) and do not necessarily reflect the views of the Under Secretary of Defense for Research and Engineering or the US Government.

L.D., M.H., and Y.S. performed the experiments and analyzed the data.
L.D., M.H., and J.A. developed the theory and numerical simulations.
L.D., Y.S., A.D.P., and K.S. designed the device.
K.A., D.K.K., B.M.N., A.M., M.E.S., and J.L.Y. fabricated and packaged the devices. 
L.D., M.H., Y.S., B.K., J.A., A.H.K., and S.G. contributed to the experimental setup.
T.H., T.P.O., S.G., J.A.G., K.S., and W.D.O supervised the project.
L.D. and M.H. wrote the manuscript with feedback from all authors. 
All authors contributed to the discussion of the results and the manuscript. 

\end{acknowledgments}

\clearpage

\appendix

\section{Wiring}

\begin{figure}[!htb]
\includegraphics{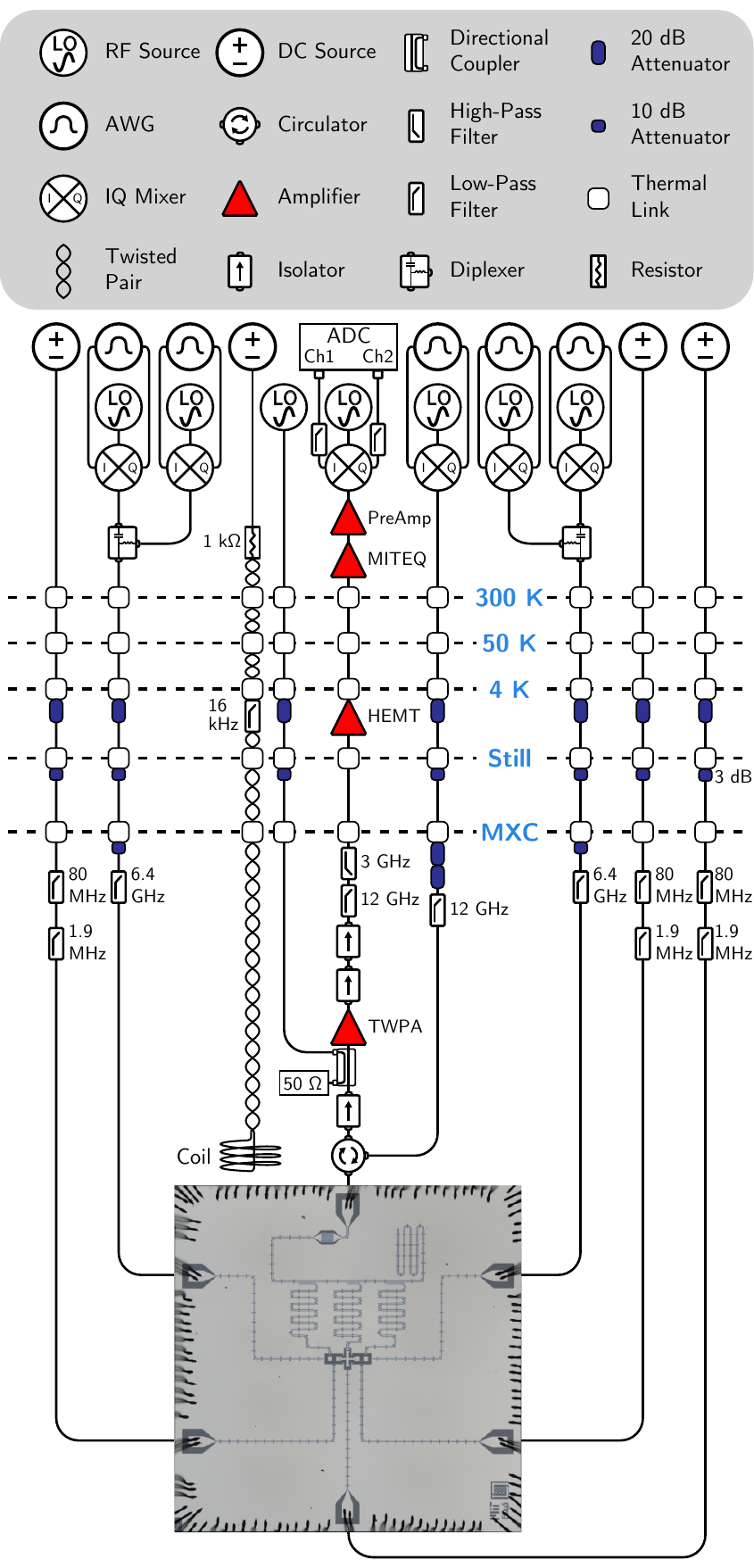}
\caption{\label{fig:sup_wiring} \textbf{A detailed wiring schematic of the experimental setup.}} 
\end{figure}

This experiment was conducted in a Bluefors XLD600 dilution refrigerator operated at around \SI{20}{mK}, with the full wiring setup shown in \fref{fig:sup_wiring}. At the mixing chamber (MXC), the device was magnetically shielded with a superconducting can, surrounded by a Cryoperm-10 can. To reduce thermal noise from higher temperature stages, we typically used in total \SI{23}{dB} attenuation for the coupler flux lines, \SI{30}{dB} attenuation for the fluxonium flux lines, \SI{40}{dB} total attenuation on charge lines, and \SI{70}{dB} total attenuation on the readout input -- the exact value of the attenuation at Still varied between \SI{3}{dB} and \SI{10}{dB} across the flux lines of both devices, though this difference was not critical for any experiment. The readout output was first amplified by a Josephson traveling-wave parametric amplified (JTWPA), pumped by a Holzworth RF synthesizer, then amplified further with a high-electron-mobility transistor (HEMT) amplifier at the \SI{4}{K} stage, another HEMT at room temperature, and a final Stanford Research SR445A amplifier, before being digitized by a Keysight M3102A digitizer. 

All AC signals -- readout, single- and two-qubit gate pulses -- were generated by single sideband mixing of Keysight M3202A 1GSa/s arbitrary waveform generators with Rohde and Schwarz SGS100A SGMA RF sources. For each qubit, the single- and two-qubit gate pulses were combined at room temperature via a diplexer from Marki Microwave (DPXN-2 for Qubit 1 and DPXN-0R5 for Qubit 2). For these diplexers, the single-qubit gate frequencies occur at low enough frequencies to fall in the pass band of the DC port. All control electronics were synchronized through a common SRS \SI{10}{MHz} rubidium clock. 

The DC voltage bias for each qubit flux line as well as the global bobbin was supplied by a QDevil QDAC. The flux lines by design support RF flux, but in this experiment were filtered by \SI{80}{MHz} and \SI{1.9}{MHz} low-pass filters at the MXC.  The current for the global coil was carried through a twisted pair, with a homemade \SI{16}{kHz} cutoff RC filter at the \SI{4}{K} stage. 

\begin{table}[!htb]
\caption{\label{tab:equipment}
\textbf{Summary of control equipment.} The manufacturers and model numbers of the control equipment used for this experiment.
}
\begin{tabular}{|c|c|c|}
\hline
Component & Manufacturer & Model\\
\hline
Dilution Fridge & Bluefors & XLD600\\
RF Source & Rohde and Schwarz & SGS100A\\
DC Source & QDevil & QDAC I\\
Control Chassis & Keysight & M9019A\\
AWG & Keysight & M3202A\\
ADC & Keysight & M3102A\\
\hline
\end{tabular}
\end{table}

\section{Grounded or differential qubits?} \label{appendix:grounded_vs_floating}

In our circuit design, we made a conscious decision on whether each qubit (or coupler) should be made a grounded qubit or a differential qubit. By using a differential fluxonium, we reduce the amount of capacitance that coupling appendages contribute to the total effective qubit capacitance. A differential qubit also allows for a larger total area of capacitor pads for the same qubit charging energy~$E_C$. This is important to allow for enough physical room to couple other circuit elements such as resonators, charge lines, flux lines, and other qubits to each fluxonium. 

\subsection{Grounded transmon}
\begin{figure}[!htb]
\includegraphics{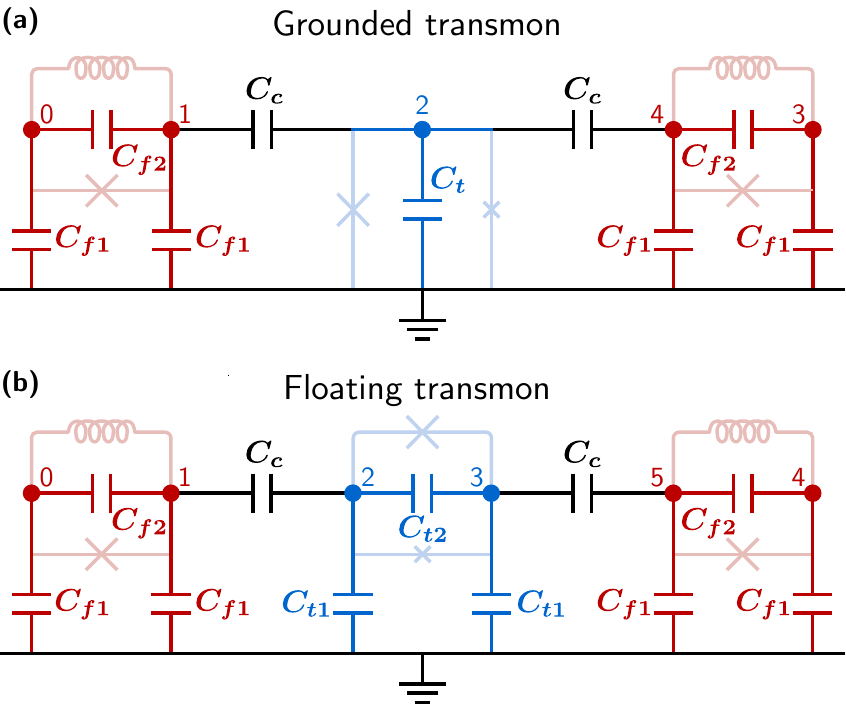}
\caption{\label{fig:sup_cap_matrix} \textbf{Simplified circuit model of FTF circuits.} \textbf{(a)} FTF circuit with a grounded transmon. The capacitance network is simplified for the purpose of a theoretical analysis, with no direct fluxonium-fluxonium capacitance. \textbf{(b)} Same circuit except with a differential transmon coupler.} 
\end{figure}

The choice to use a grounded transmon, on the other hand, stems from the relationship between~$J_c=J_{1c}=J_{2c}$ and~$J_{12}$. As mentioned in the main text, to minimize the~$ZZ$ interaction~$\zeta$, we need~$J_c^2 / J_{12} = - 2 \zeta^{(2)} / \zeta^{(3)} \approx \SI{2.97}{GHz}$ for our device parameters. We claim that by using a grounded transmon, we can target this value of~$J_c/J_{12}$ with first-order insensitivity to the coupling capacitance (between the transmon and the adjacent fluxonium pad)~$C_c$. Consequently, $C_c$ becomes a free parameter in the device design, and uncertainty in its value will, to first order, have no effect on~$\zeta$. To understand this theoretically, we assume the simplified circuit schematic represented in \fref{fig:sup_cap_matrix}\panel{a}. All capacitances not explicitly labeled are small and qualitatively unimportant in this analysis. 

\noindent We can write down the capacitance matrix of this circuit as 
\begin{equation}
\mathbf{C} = \begin{pmatrix}
	C_{F} & - C_{f2} & 0 & 0 & 0\\
	- C_{f2} & C_{F} + C_c  & -C_c & 0 & 0\\
	0 & -C_c & C_t + 2 C_c & 0 & -C_c\\
	0 & 0 & 0 & C_{F} & -C_{f2}\\
	0 & 0 & -C_c & -C_{f2}  & C_{F} + C_c
\end{pmatrix},\end{equation}
where we defined~$C_F = C_{f1} + C_{f2}$ for convenience. In order to isolate the relevant mode of each differential qubit, we perform a standard variable transformation into sum and difference coordinates which modifies the capacitance matrix as~$\tilde{\mathbf{C}} = (\mathbf{M}^T)^{-1} \mathbf{C} \mathbf{M}^{-1}$ with
\begin{equation}
\mathbf{M} = \begin{pmatrix}
1 & 1 & 0 & 0 & 0\\
1 & -1 & 0 & 0 & 0\\
0 & 0 & 1 & 0 & 0\\
0 & 0 & 0 & 1 & 1\\
0 & 0 & 0 & 1 & -1
\end{pmatrix}.
\end{equation}
The qubit modes for each differential qubit are then solely determined by the difference coordinates, with the resultant three qubit nodes on indices 1, 2, and 4 (counting from 0). We can straightforwardly discard the modes corresponding to summed coordinates in the Hamiltonian and compute coupling strengths between nodes as~$J_{ij} = 4e^2 \tilde{\mathbf{C}}^{-1}[i, j]$. Thus,
\begin{align}
    J_c^2/J_{12} &= 4e^2 \frac{\tilde{\mathbf{C}}^{-1}[1, 2]^2}{\tilde{\mathbf{C}}^{-1}[1, 4]} \label{eq:ratio_grounded_exact}\\
    &\approx 4e^2 \frac{1}{C_t} + \mathcal{O}(C_t^{-2}), \label{eq:ratio_grounded}
\end{align}
where we performed a Taylor expansion assuming~$C_c, C_{f1}$, and~$C_{f2}$ are small compared to~$C_t$ in the final step. We see that to leading order, the value of~$J_c^2/J_{12}$ is solely determined by~$C_t^{-1}$, with any dependence on~$C_c$ scaling with~$\mathcal{O}(C_t^{-2})$. By inserting the designed values of~$C_t = \SI{45}{fF}$ ~$C_{f1} = \SI{11}{fF}$, and~$C_{f2} = \SI{2.8}{fF}$, \eref{eq:ratio_grounded_exact} gives \SI{2.8}{GHz} and \eref{eq:ratio_grounded} gives \SI{3.4}{GHz}. We emphasize that \eref{eq:ratio_grounded} illustrates a concept in our architecture and that exact design simulations of our coupling strengths were performed using full 5$\times$5 capacitance matrices with no mathematical approximations.

\subsection{Differential transmon}
To investigate how these relationships would compare when substituting for a differential transmon, we model the hypothetical circuit with the capacitance network in \fref{fig:sup_cap_matrix}\panel{b}. The capacitance matrix, in this case, is
\begin{align}
&\mathbf{C} = \notag \\
&\begin{pmatrix} 
C_{F} & - C_{f2} & 0 & 0 & 0 & 0\\
- C_{f2} & C_{F} + C_c  & -C_c & 0 & 0 & 0\\
0 & -C_c & C_{T} + C_c & -C_{t2} & 0 & 0\\
0 & 0 & -C_{t2} & C_{T} + C_c & 0 & -C_c\\
0 & 0 & 0 & 0 & C_{F} & -C_{f2}\\
0 & 0 & 0 & -C_c & -C_{f2}  & C_{F} + C_c
\end{pmatrix}
\end{align}
where we have likewise defined~$C_T = C_{t1} + C_{t2}$. The transformation matrix in this case is 
\begin{equation}
\mathbf{M} = \begin{pmatrix}
1 & 1 & 0 & 0 & 0 & 0\\
1 & -1 & 0 & 0 & 0 & 0\\
0 & 0 & 1 & 1 & 0 & 0\\
0 & 0 & 1 & -1 & 0 & 0\\
0 & 0 & 0 & 0 & 1 & 1\\
0 & 0 & 0 & 0 & 1 & -1
\end{pmatrix}
\end{equation}
and our coupling ratio is
\begin{align}
    J_c^2/J_{12} &= 4e^2 \frac{\tilde{\mathbf{C}}^{-1}[1, 3]^2}{\tilde{\mathbf{C}}^{-1}[1, 5]} \label{eq:ratio_floating_exact}\\
    &\approx 4e^2 \frac{1}{C_{t2}} + \mathcal{O}(C_{t1}^{-1}). \label{eq:ratio_floating}
\end{align}
While still independent of~$C_c$ to leading order, the value of~$4e^2/C_{t2}$ is far too large compared to optimal values. Furthermore, FTF benefits from as high of coupling strengths as possible, and a differential transmon reduces the values of~$J_c$ and~$J_{12}$ for a fixed value of~$C_c$. 

\section{\textit{ZZ} cancellation in FTF} \label{appendix:ZZ}
In this section, we elaborate on the theory which gives rise to the reduced $ZZ$ rate in the FTF architecture. We specifically consider the level repulsion on the $\ket{101}$ state, as we find it most impacted by the coupling between qubits, and as a result, indicative of how level repulsions affect the $ZZ$ rate as a whole. We will consider the energy shift of the $\ket{101}$ to up to 4th order in perturbation theory, where $n$ represents the $\ket{101}$ state, $k_i$ represent any intermediate state that is not $n$, $V_{jk}$ represents the Hamiltonian matrix element $\bra{j}H\ket{k}$ between the two bare states $\ket{j}$, $\ket{k}$, and $E^{(m)}_n$ represents the contribution to the energy of the state $\ket{n}$ at perturbative order $m$. Energy detunings between two states $\ket{j}$ and $\ket{k}$ are similarly denoted $E^{(m)}_{jk} = E^{(m)}_j - E^{(m)}_k$. In this notation, the 2nd order correction to the energy of $\ket{n}$ is 
\begin{equation} \label{eq:pt2}
    E_n^{(2)} = \sum_{k_1} \frac{|V_{n k_1}|^2}{E_{n k_1}^{(0)}}.
\end{equation}
Due to the relatively high energy of the fluxonium $\ket{2}$ states and the transmon $\ket{1}$ state, all intermediate states in this summation have higher energy than $\ket{101}$. The only exception is the $\ket{000}$ state; however, any energy shift involving only states in the computational basis will not contribute to the total $ZZ$ rate because the equal and opposite level repulsions will cancel out when computing $\zeta$. As a result, $E_n^{(2)}$ is negative and independent of the coupler element, provided the computational states are formed by capacitively coupled fluxonium qubits. 

To fourth order in perturbation theory, the energy shift is
\begin{equation} \label{eq:pt4}
    E_n^{(4)} = \sum_{k_1 k_2 k_3} \frac{V_{n k_1} V_{k_1 k_2} V_{k_2 k_3} V_{k_3 n}}{E_{n k_1}^{(0)} E_{n k_2}^{(0)} E_{n k_3}^{(0)}} - E_n^{(2)} \sum_{k_1} \left|\frac{V_{n k_1}}{E_{n k_1}^{(0)}}\right|^2.
\end{equation}
While the expression here is more complicated, the qualitative result is the same. Since each energy detuning in the denominators is negative when all intermediate states have higher energy than $\ket{n}$, the resultant quantity $E_n^{(4)}$ is also negative. 

\begin{figure}[!htb]
\includegraphics{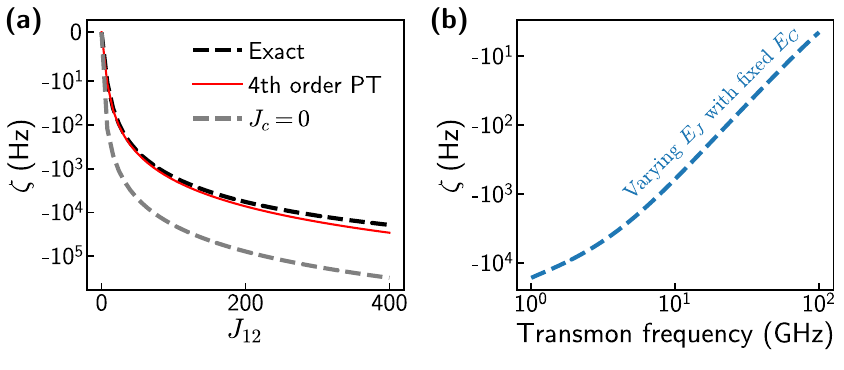}
\caption{\label{fig:sup_zz} \textbf{Numerical simulations of $\zeta$ in the FTF system.} \cpanel{a} Plot of $\zeta$ along the `Minimum $ZZ$' parabola in \fref{fig:zz}\panel{a}, as a function of $J_{12}$. The black curve shows a numerical diagonalization of the Hamiltonian, which is accurately described by 4th order perturbation theory (red).  Without the coupler (gray), $|\zeta|$ is roughly an order of magnitude larger for this range of coupling strengths. \cpanel{b} Numerical simulation of $\zeta$ as a function of the coupler frequency. The Device A parameters from \tref{tab:hamiltonian} are used, except the effective transmon $E_J$ is changed to vary the transmon frequency. As this frequency increases, $|\zeta|$ decreases asymptotically to below \SI{10}{Hz} for this set of device parameters.}
\end{figure}

Unlike the second and fourth order terms, third order contributions can only exist when both fluxonium qubits are coupled to a third qubit
\begin{equation} \label{eq:pt3}
    E_n^{(3)} = \sum_{k_1 k_2} \frac{V_{n k_1} V_{k_1 k_2} V_{k_2 n}}{E_{n k_1}^{(0)} E_{n k_2}^{(0)}}.
\end{equation}
Contrary to the even-order terms, all third-order contributions are positive, providing the critical mechanism to reduce the $ZZ$ rate. While this analysis has been done just for the energy of the $\ket{101}$ state, numerical computations of \eref{eq:pt2}, \eref{eq:pt4}, and \eref{eq:pt3} confirm that second and fourth order terms contribute negatively to the overall $ZZ$, and the third order terms contribute positively to the overall $ZZ$. In \fref{fig:sup_zz}\panel{a}, we compute $\zeta$ as a function of $J_{12}$ (assuming the optimal corresponding value of $J_c$) with an exact numerical diagonalization (black) and the aforementioned perturbation theory (red), illustrating that perturbation theory up to 4th order is a sufficient description of the total $ZZ$ rate in the FTF system. 

Another interesting feature of FTF unexplored in the main text is that if the transmon frequency is tuned toward infinity, $\zeta$ does not return to its value with only the two fluxonium qubits and can in fact decrease even further. In \fref{fig:sup_zz}\panel{b}, we numerically simulate  $\zeta$ as a function of the coupler frequency using the Device A parameters and find that $\zeta$ asymptotes to less than \SI{10}{Hz} as the transmon frequency increases. When the transmon frequency is tuned, not only do the energy level detunings increase, but the charge matrix element also increases through the effective $E_J$. This increase in the charge matrix elements prevents the level repulsions involving the coupler from vanishing even when the coupler frequency becomes infinitely large. This feature makes FTF useful even in fluxonium gate schemes that do not involve the non-computational states. By coupling a fixed frequency transmon (or resonator) with a high frequency (arising from a high $E_J$) to each fluxonium with the appropriate coupling strengths, $\zeta$ can be reduced to near 0 in a robust manner without needing any measurement or calibration of the extra transmon element. 

\section{Readout} \label{appendix:readout}

\begin{figure}[!htb]
\includegraphics{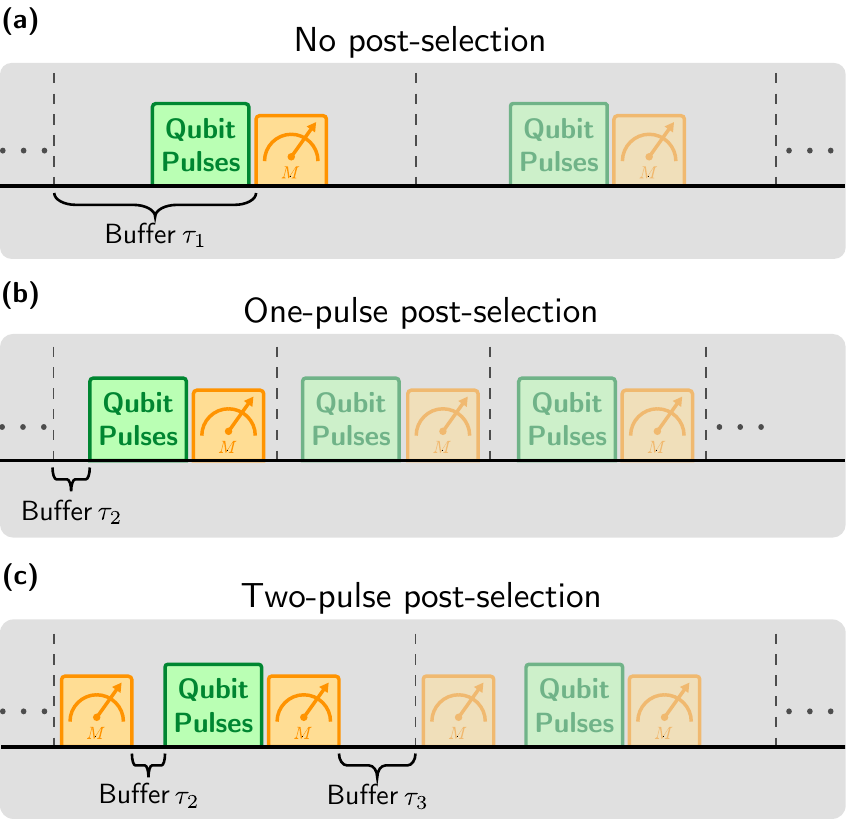}
\caption{\label{fig:sup_readout} \textbf{Pulse diagram for different readout configurations.} \cpanel{a} Standard measurement and readout sequence. Pulses are played on a repeated trigger so that the time between the trigger start and the readout start is kept constant. \cpanel{b} Single-readout post-selection sequence, also known as heralding. Each readout pulse simultaneously sets the initial state for the subsequent qubit pulses and records the measurement outcome of the previous qubit pulses. To herald the fluxonium ground (excited) state~$\ket{0}$ ($\ket{1}$), we only accept measurement results for which the previous readout result was~$\ket{0}$ ($\ket{1}$). \cpanel{c} Two-pulse post-selection (heralding) technique. The first readout is used to initialize the qubit state, and the second is used to measure the result of the qubit pulses. This extra readout pulse allows for an additional buffer time~$\tau_3$ without impacting the fidelity of the state preparation.} 
\end{figure}

In our standard readout pulse-sequence [\fref{fig:sup_readout}\panel{a}], the qubit and readout pulses were periodically played corresponding to a software trigger, with a fixed wait time~$\tau_1$ between the start of the trigger and the start of the readout pulses. The trigger period and~$\tau_1$ were chosen such the qubit sufficiently decays to its thermal equilibrium state between measurements.

To initialize the fluxonium qubits in either $\ket{0}$ or $\ket{1}$, we relied on a mostly QND single-shot readout. Then, we post-selected the data for any choice of initial state:~$\{\ket{0}, \ket{1}\} \otimes \ket{0} \otimes \{\ket{0}, \ket{1}\}$ in order to herald that state~\cite{Johnson2012}.
In our simplest form of post-selection [\fref{fig:sup_readout}\panel{b}], the qubit and readout pulses were played back-to-back with only a short buffer time~$\tau_2$ to allow for photons to depopulate the readout resonator prior to the qubit pulses. In our experiment,~$\tau_2 = \SI{2}{\micro s}$ was much less than the~$T_1$ of any qubit, so the qubits do not return to their thermal equilibrium state by the start of the next pulse sequence. Despite losing a fraction of our data due to this post-selection process, we gain an enormous overhead by not having to wait for the qubits to decay between measurements. Using this method, we obtained typical initialization fidelities between 95\% and 99\% for~$\ket{0}$ and~$\ket{1}$ on each qubit. This was calculated as the probability of measuring~$\ket{0}$ ($\ket{1}$) if the previous measurement result was also~$\ket{0}$ ($\ket{1}$). Due to previously documented non-QND readout effects which transfer population from~$\ket{0}$ to~$\ket{1}$~\cite{Ficheux2021}, the initialization fidelity of the ground state was often 1-2\% worse than the excited state.

Ideally, the above method of post-selection would work with the higher-excited states of the fluxonium, but we found in practice that we could not distinguish the~$\ket{1}$ and~$\ket{2}$ states of the fluxonium in single-shot. To work around this issue, we used a separate two-pulse post-selection technique [\fref{fig:sup_readout}\panel{c}]. The first readout initializes the qubit in either~$\ket{0}$ or~$\ket{1}$, and the second readout records the measurement result. We used the same buffer time~$\tau_2$ to avoid measurement-induced dephasing of our qubit and introduced an additional wait time~$\tau_3 \sim \SI{50}{\micro s}$ to let any population in the non-computational states of the fluxonium to decay to the computational states.
Many of our measurements specifically involved driving from~$\ket{1}$ to~$\ket{2}$ on a particular fluxonium (we include transitions such as~$\ket{101} \leftrightarrow \ket{111}$, $\ket{101} \leftrightarrow \ket{102}$, and~$\ket{101} \leftrightarrow \ket{201}$ in this discussion). In these cases, we could greatly enhance the readout contrast by performing a~$\pi$-pulse on each fluxonium prior to readout. After this~$\pi$-pulse, any population measured in~$\ket{0}$ was assumed to be originally~$\ket{1}$, and any population measured in~$\ket{1}$ was assumed to be originally~$\ket{2}$.   

\section{Single-qubit gate calibration} \label{appendix:single_qubit}

Our single-qubit gates were performed using standard Rabi oscillations using a cosine envelope without a flat top. For the measurements in \fref{fig:1qb}, we included a \SI{4}{ns} zero-padding between adjacent pulses. We individually calibrated the~$X_\pi$ pulse and derived all other pulses from it. $Y$ pulses were created by adjusting the phase of the~$X$ pulses, pulses differing from~$\pi$ rotation were derived by linearly scaling the pulse amplitude, and~$Z$ gates were all implemented as virtual-Z gates. While the set of single-qubit gates could be more carefully calibrated by individually calibrating all other gates, we found our single-qubit gates derived from this process were more than sufficient for accurately calibrating our CZ gate.

\begin{figure*}[!htb]
\includegraphics{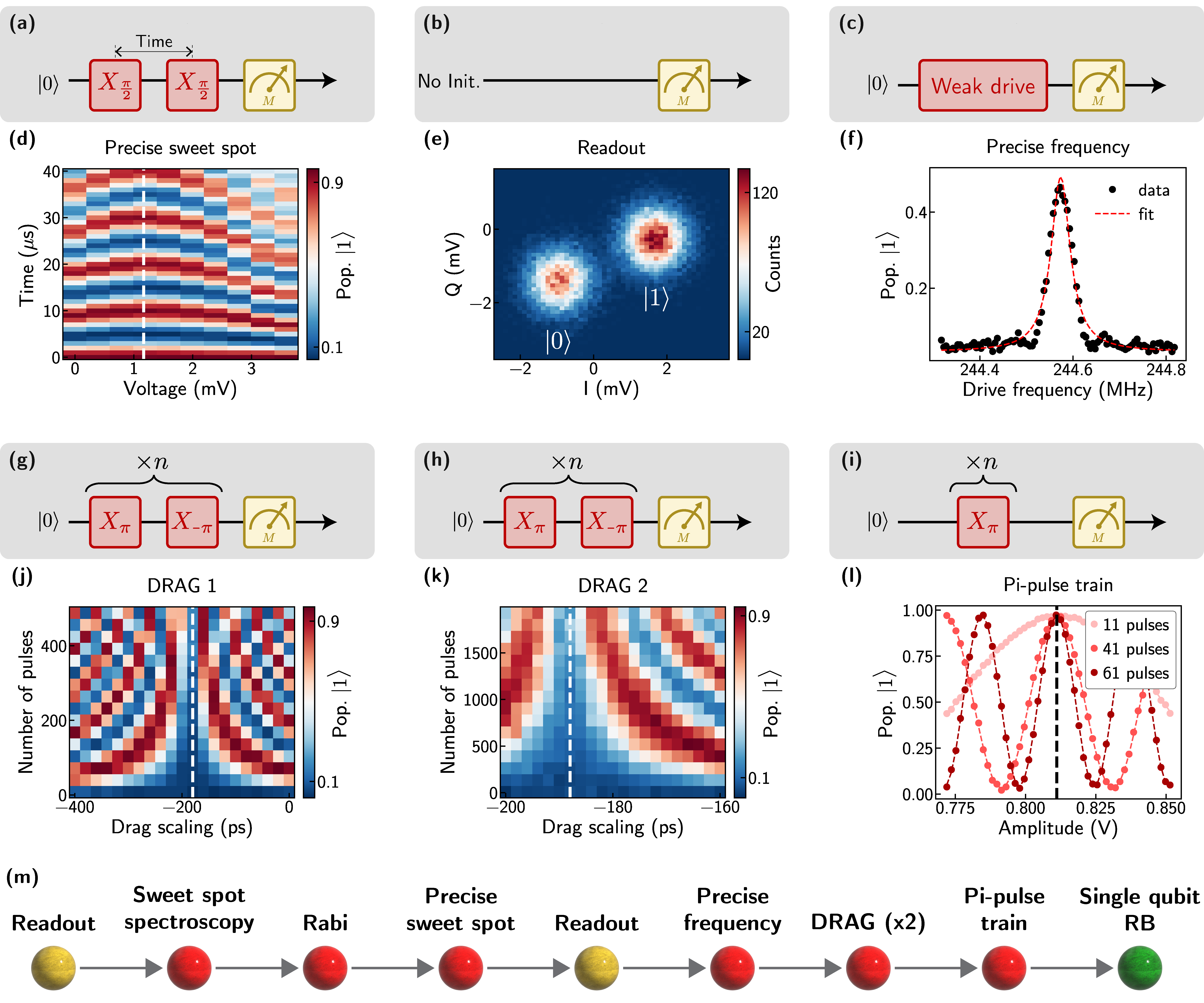}
\caption{\label{fig:sup_1qb_cal} \textbf{Single-qubit~$X_\pi$ calibration procedure.} \cpanel{a-c, g-i} Measurement pulse sequences for panels \cpanel{d-f, j-l} respectively. \cpanel{d} Ramsey vs. bias voltage measurement in order to precisely determine the voltage corresponding to~$\Phi_\text{ext} = \SI{0.5}{\Phi_0}$ (termed the `sweet spot'). \cpanel{e} Calibrating the readout I and Q coordinates for the fluxonium~$\ket{0}$ and~$\ket{1}$ states. Thermal population of the qubit results in a mixed state prior to readout. \cpanel{f} Low power spectroscopy of the qubit to precisely determine the qubit frequency. \cpanel{j} DRAG calibration of the qubit. We played a varying even number of~$X_\pi$ pulses with alternating amplitude while also varying the DRAG parameter. When correctly calibrated, each~$X_\pi$ pulse cancels out with the subsequent~$X_{-\pi}$ pulse. \cpanel{k} A separate DRAG calibration but using a larger number of pulses to calibrate the DRAG parameter more accurately. \cpanel{l} A pulse train consisting of an odd number of~$X_\pi$ pulses was used to precisely calibrate the amplitude of the pulse. \cpanel{m} Calibration flowchart for single-qubit gates. Panels are shown for all calibrations starting from the `Precise sweet spot' calibration.} 
\end{figure*}

We detail the entire calibration sequence used for the~$X_\pi$ gate in \fref{fig:sup_1qb_cal}, with a flowchart provided in \fref{fig:sup_1qb_cal}\panel{m}. Following an initial rough calibration of the qubit readout, bias voltage, frequency, and~$\pi$-pulse amplitude, the following procedure was used to precisely calibrate the qubit.

\begin{enumerate}
    \item \textbf{Precise sweet-spot calibration} [\fsref{fig:sup_1qb_cal}\panel{a,~d}]. With fixed drive frequency (slightly negatively detuned from the qubit frequency) and fixed drive amplitude, Ramsey oscillations were obtained as a function of flux around the sweet spot. The oscillation frequencies were fit to a parabola, and the center of the parabola fit was used as the bias voltage corresponding to~$\Phi_\text{ext} = \SI{0.5}{\Phi_0}$, termed the `sweet spot'. 
    
    \item \textbf{Single-shot readout calibration} [\fsref{fig:sup_1qb_cal}\panel{b,~e}]. Although optimizing the readout does not impact the gate fidelity, we re-measured the locations of the single-shot blobs corresponding to~$\ket{0}$ and~$\ket{1}$ here to correct for flux-related changes in the readout signal.
    
    \item \textbf{Precise qubit frequency calibration} [\fsref{fig:sup_1qb_cal}\panel{c,~f}]. To accurately obtain the qubit frequency, we performed qubit spectroscopy with a low enough power such that little power broadening was observed. This typically gave kHz-level precision, a sufficient starting point for DRAG. 
    
    \item \textbf{Derivative Removal by Adiabatic Gate (DRAG) calibration} [\fsref{fig:sup_1qb_cal}\panel{g-h,~j-k}]. While scanning the DRAG parameter, we performed a train of~$\pi$-pulses with alternating positive and negative amplitudes. The DRAG parameter which minimized the observed oscillation between the~$\ket{0}$ and~$\ket{1}$ state was chosen. This measurement may be repeated with a larger number of pulses to increase the resolution of the DRAG parameter. Contrary to other fluxonium experiments~\cite{Bao2022, Somoroff2021}, we found DRAG calibration necessary to avoid coherent additional errors. We measured the optimal DRAG parameter to vary depending on the room temperature filtering scheme and not with the anharmonicity of the qubit, which leads us to suspect the calibration was correcting for small distortions in the drive line~\cite{Gustavsson2013}. 
    
    \item \textbf{Precise drive amplitude calibration} [\fsref{fig:sup_1qb_cal}\panel{i,~l}]. To improve the precision of a single Rabi oscillation, we utilized a pulse train with an odd number of pulses while varying the pulse amplitude. This serves to multiply the oscillation frequency by the number of pulses used, allowing for increased precision on the calibrated~$\pi$-pulse amplitude.
\end{enumerate}

After calibration, we benchmarked our single-qubit gates using standard single-qubit randomized benchmarking~\cite{Magesan2011, Barends2014}. In our randomized benchmarking, we applied a sequence of random Clifford gates (decomposed into microwave gates~${\{I,\, \pm X,\, \pm Y,\, \pm X_{\pi/2},\, \pm Y_{\pi/2}\}}$) followed by a recovery gate which inverts the previous sequence and then measure the probability the qubit remains in its initial state (we arbitrarily choose~$\ket{0}$). As a function of Clifford sequence length, this probability was fit to the exponential decay:~${p_0(m) = A p^m + B}$. In this model, $A$ and~$B$ absorb the effects of state preparation and measurement (SPAM) errors, and~$p$ is the depolarizing parameter. The average fidelity of each Clifford operation is given by~$F_\text{Clifford} = 1 - (1 - 1/d)(1 - p)$, where~$d=2^\text{number of qubits}$ is the size of the Hilbert space. Our specific decomposition of Clifford gates into microwave single-qubit gates uses, on average, 1.875 gates per Clifford. From this, we compute the average single-qubit gate fidelity as
\begin{equation}
F_\text{single-qubit} = 1 - (1 - 1/d)(1 - p) / 1.875. 
\end{equation}
The uncertainty on the fidelity is expressed as a standard error of the mean, which was obtained by setting \texttt{absolute\_sigma} to be \texttt{True} in \texttt{scipy.optimize.curve\_fit} and using standard error propagation techniques. Error bars on individual RB points are likewise standard errors, obtained by dividing the standard deviation by the square root of the number of randomizations.

\section{Leakage transitions} \label{appendix:transitions}

\begin{figure*}[!htb]
\includegraphics{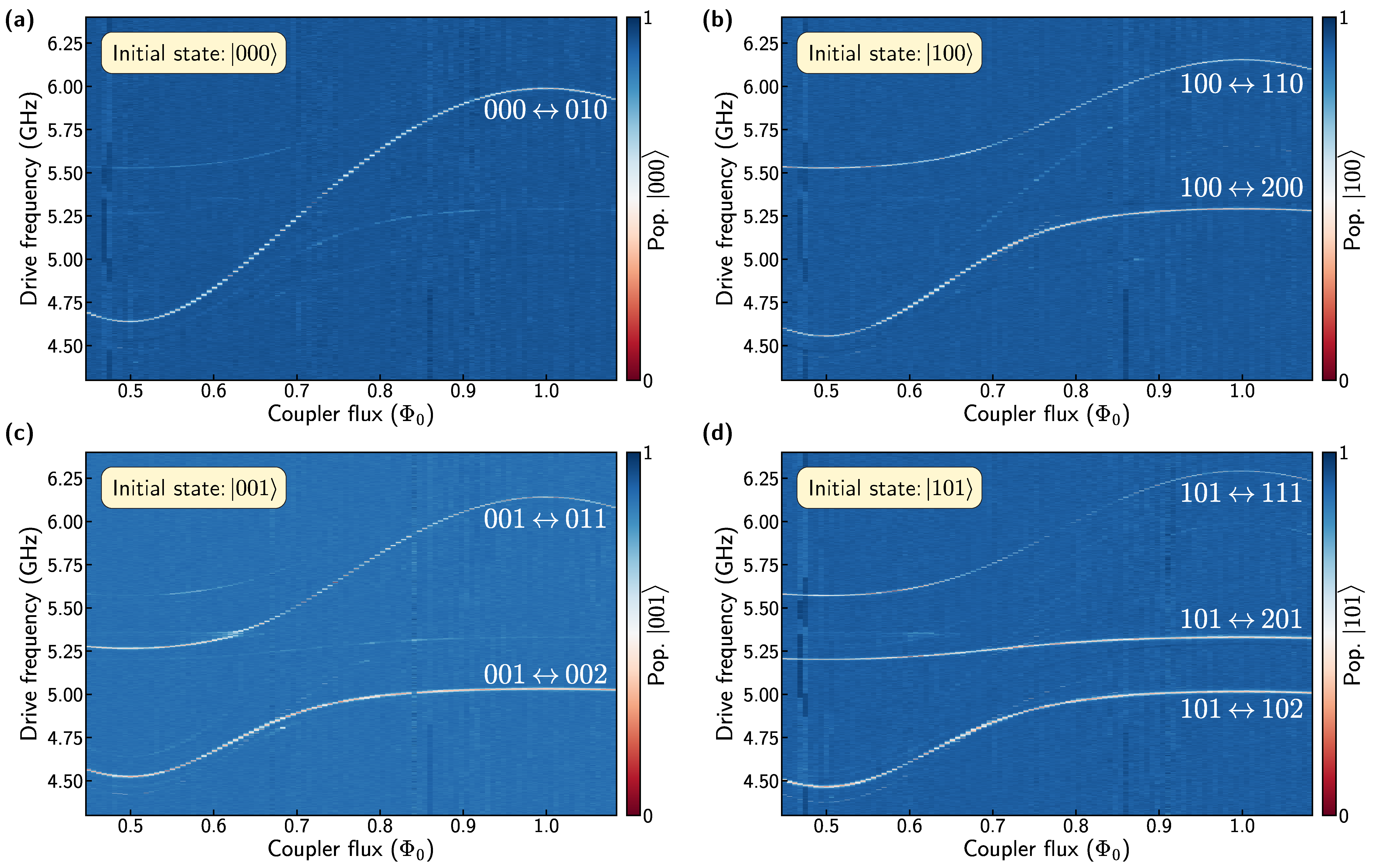}
\caption{\label{fig:sup_sidebands}
\textbf{Two-tone spectroscopy of the higher energy levels of the FTF system.} Panels \cpanel{a-d} consist of the same spectroscopy run post-selecting for~$\ket{000}$, $\ket{100}$, $\ket{001}$, or~$\ket{101}$ respectively.}
\end{figure*}

Driving a single transition in a system of uncoupled qubits is generally equivalent to a single-qubit rotation of some kind. For our CZ gate, such a transition is only entangling provided no other transitions from other initial states are also being driven. This is only possible when coupling terms adjust the energy levels involved in these transitions. To map out the landscape of the relevant higher-level transitions, we performed two-tone spectroscopy as a function of the coupler flux, heralding different initial states:~$\ket{000}$, $\ket{100}$, $\ket{001}$, and~$\ket{101}$ [\fref{fig:sup_sidebands}]. As demonstrated in the main text, any transition in \fref{fig:sup_sidebands}\panel{d} could be used to perform the CZ gate, provided it is sufficiently detuned from any transitions involving the other initial states. Equivalently, though not explored in this work, a CZ gate could be performed by driving selective transitions from a different initial state and then performing appropriate~$Z$ rotations to transfer the~$180^\circ$ conditional phase shift onto the~$\ket{11}$ state. 

While including a coupler increases the number of parasitic transitions that must be avoided, the larger level repulsions made possible by the coupler increase many relevant detunings. In our devices, many operational points exist where the nearest unwanted transition is more than \SI{100}{MHz} detuned. 

\section{Relative drive amplitude and phase calibration}

In this section, we describe a leakage cancellation protocol utilizing destructive interference of the two drives; however, we note that a difference in gate fidelity could not be observed using this method. As a result, for the data in \fref{fig:map_spec}\panel{c}, a simpler procedure was implemented to save calibration time: the relative drive phase was tuned to constructively interfere at the desired transition, and the relative amplitudes were kept equal. Despite this fact, we include the information here for transparency on what was attempted to improve gate fidelities.

In driving our desired transition, off-resonant parasitic transitions always contribute to leakage. With two separate charge drive lines in our device, we can tune each drive line's relative phase and amplitudes for complete destructive interference on a parasitic transition of our choosing, while retaining a nonzero drive on our gate transition. Without loss of generality, we take~$\ket{101} \leftrightarrow \ket{111}$ to be our CZ gate transition and~$\ket{100} \leftrightarrow \ket{200}$ to be the closest parasitic transition that we'd like to eliminate. We model the pulse seen by the qubits~$i \in \{1, 2\}$ as 

\begin{equation}
    \text{Pulse}_i(t) = A_i \cos(\omega t + k x_i + \phi_i),
\end{equation}
where~$A_i > 0$ is the pulse amplitude,~$\omega$ is the angular frequency,~$k$ is the wavenumber of the pulse,~$x_i$ is the effective distance from each pulse's origin to its destination, and~$\phi_i$ is an additional constant phase offset of each pulse, specified in software. 

The Rabi frequency of the undesired transition can then be written as 
\begin{equation}
    \bra{200}\hat{H}\ket{100} \propto \sum_i A_i \cos(\omega t + k x_i + \phi_i) \bra{200}\hat{n}_i\ket{100} 
\end{equation}
For this matrix element to be zero for all times, we require the two pulses to be~$180^\circ$ out of phase with each other with equal effective amplitudes. Mathematically, these two conditions are satisfied by specifying the relative phase and amplitude of the two drives:
\begin{align}
    \phi_2 - \phi_1 &= \pi - k(x_2 - x_1)\label{eq:phase} \\
    \frac{A_2}{A_1} &= \frac{\bra{200}{\hat n_1}\ket{100}}{\bra{200}{\hat n_2}\ket{100}}, \label{eq:amplitude}
\end{align}
We note that since~$\bra{200}\hat n_1\ket{100}/\bra{200}\hat n_2\ket{100}$ is generally not equal to~$\bra{111}\hat n_1\ket{101}/\bra{111}\hat n_2\ket{101}$, these conditions are not expected to provide complete destructive interference on our main transition of interest. 

\begin{figure}[!htb]
\includegraphics{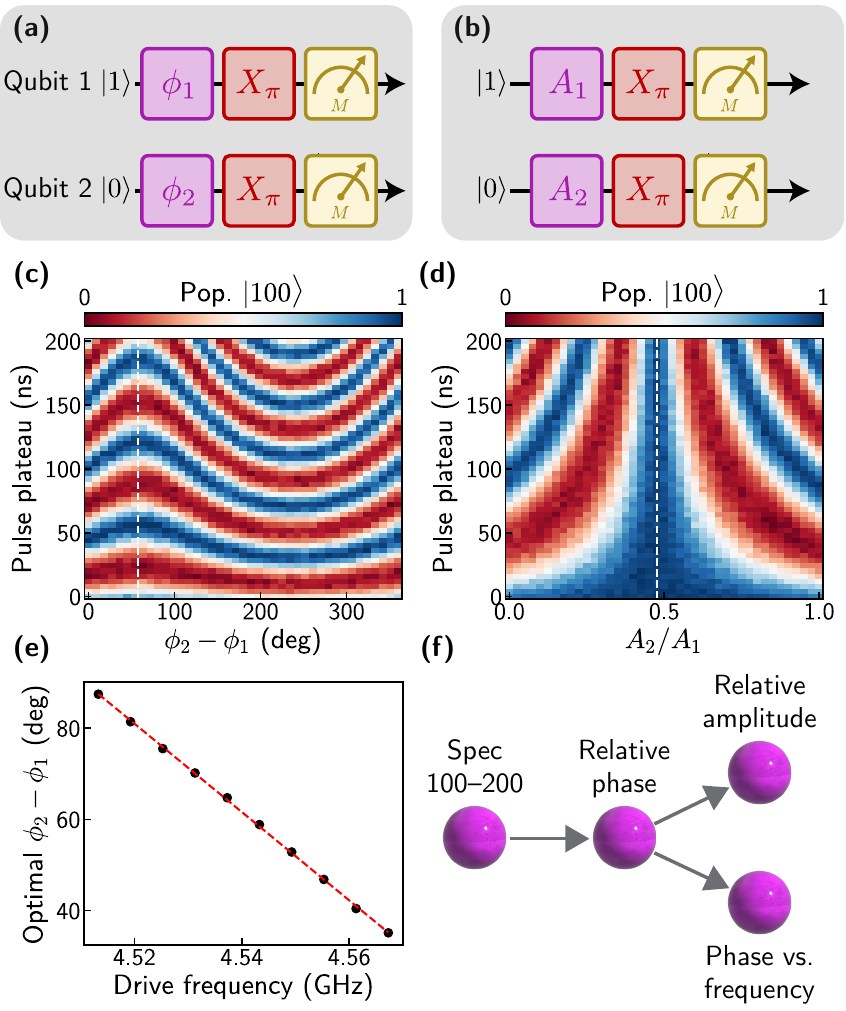}
\caption{\label{fig:sup_microwave_crosstalk} \textbf{Relative drive amplitude and phase calibration.} \cpanel{a, b} Pulse sequences for (d-f) respectively.~$\pi$-pulses before measurement are purely for increasing signal contrast. \cpanel{c} Rabi oscillation of the unwanted transition with a phase-locked drive on each charge line. The phase difference was scanned to show destructive (white dashed line) and constructive interference. \cpanel{d} With~$\phi_2 - \phi_1$ specified to give destructive interference, the relative amplitudes of the two drives were scanned for \textit{complete} destructive interference (white dashed line). \cpanel{e} The value of~$\phi_2 - \phi_1$ that gave destructive interference was extracted as a function of drive frequency. The linear fit is motivated by a cable length difference. \cpanel{f} Calibration flowchart for the illustrated procedure.} 
\end{figure}

Experimentally, our procedures for calibrating~$A_2/A_1$ and~$\phi_2 - \phi_1$ are illustrated in \fref{fig:sup_microwave_crosstalk}. In all measurements for this calibration, we~$\pi$-pulsed both fluxonium qubits before readout as discussed in \aref{appendix:readout} to increase signal contrast.
We initially started with two arbitrarily amplitudes~$A_1, A_2$, and then scanned the phase difference~$\phi_2 - \phi_1$ (varying~$\phi_2$ with~$\phi_1$ fixed) while measuring the~$\ket{100} \leftrightarrow \ket{200}$ Rabi oscillation (on resonance) [\fsref{fig:sup_microwave_crosstalk}\panel{a, c}]. With the value of~$\phi_2 - \phi_1$ set to minimize the oscillation rate (dashed line),~$A_2 / A_1$ was scanned (varying~$A_2$ with~$A_1$ fixed) while measuring the same Rabi oscillation [\fsref{fig:sup_microwave_crosstalk}\panel{b, d}]. The slowest Rabi oscillation (dashed line) was then used to choose the optimal value of~$A_2 / A_1$. Furthermore, as motivated by \eref{eq:amplitude}, this ratio is independent of frequency and thus didn't require further calibration. On the other hand, we were interested in the phase which caused destructive interference when driving at our two-qubit gate frequency, not at the~$\ket{100} \leftrightarrow \ket{200}$ resonance. This required calibrating for the phase dispersion caused by cable length differences. By repeating the relative phase calibration in a frequency bandwidth in which the~$\ket{100} \leftrightarrow \ket{200}$ Rabi oscillation was still visible, we uncovered an expected linear dispersion [\fref{fig:sup_microwave_crosstalk}\panel{e}], which we extrapolated as a function of drive frequency. 

\label{appendix:microwave_crosstalk}

\section{Two-qubit gate calibration} \label{appendix:two_qubit}

Prior to a fine calibration of our CZ gate, we would first calibrate single-qubit gates (\aref{appendix:single_qubit}) and the relative drive parameters (\aref{appendix:microwave_crosstalk}). Specifically, when calibrating the CZ gate, we increased the pulse width of the single-qubit gates to \SI{50}{ns}, trading off single-qubit gate fidelities so that coherent errors would not skew our tomography pulses. We also increased the padding between pulses to \SI{10}{ns} because computational state lifetimes have only a secondary effect on our CZ gate fidelities. 

For a given pulse width and gate transition, the remaining four parameters to be calibrated are the drive frequency, overall drive amplitude, and the single-qubit phase accumulations during our gate interaction. After performing spectroscopy of the transition in interest, we iteratively fine-tuned the drive amplitude and frequency. The drive amplitude was calibrated by performing a simple Rabi oscillation and minimizing the leakage [\fsref{fig:sup_2qb_cal}\panel{a, d}]. According to our error budget outlined in \aref{appendix:error}, a single CZ pulse yields a sufficiently precise calibration. The drive frequency was calibrated by performing a Ramsey-like measurement on qubit 1 to measure its phase accrual after a pulse-train of CZ gates, depending on whether or not qubit 2 started in~$\ket{0}$ or~$\ket{1}$ [\fsref{fig:sup_2qb_cal}\panel{b, e}]. The difference in this phase accrual was tuned to be~$180^\circ$, though we note that controlled-phase gates of variable angles could also be achieved. Since adjusting the drive frequency slightly changes the amplitude corresponding to a single period and vice-versa, we alternately performed these two calibrations three times in total. In practice, we found this was sufficiently accurate and much faster than performing a two-dimensional calibration for both parameters simultaneously. Finally, once the CZ interaction was properly tuned, we measured the single-qubit phase accumulation during the CZ interaction using the same Ramsey-like measurement [\fsref{fig:sup_2qb_cal}\panel{c, f}]. These~$Z$-rotations were corrected for in software through virtual-$Z$ gates. We illustrate the complete flowchart of this calibration in \fref{fig:sup_2qb_cal}\panel{g}.

\begin{figure*}[!htb]
\includegraphics{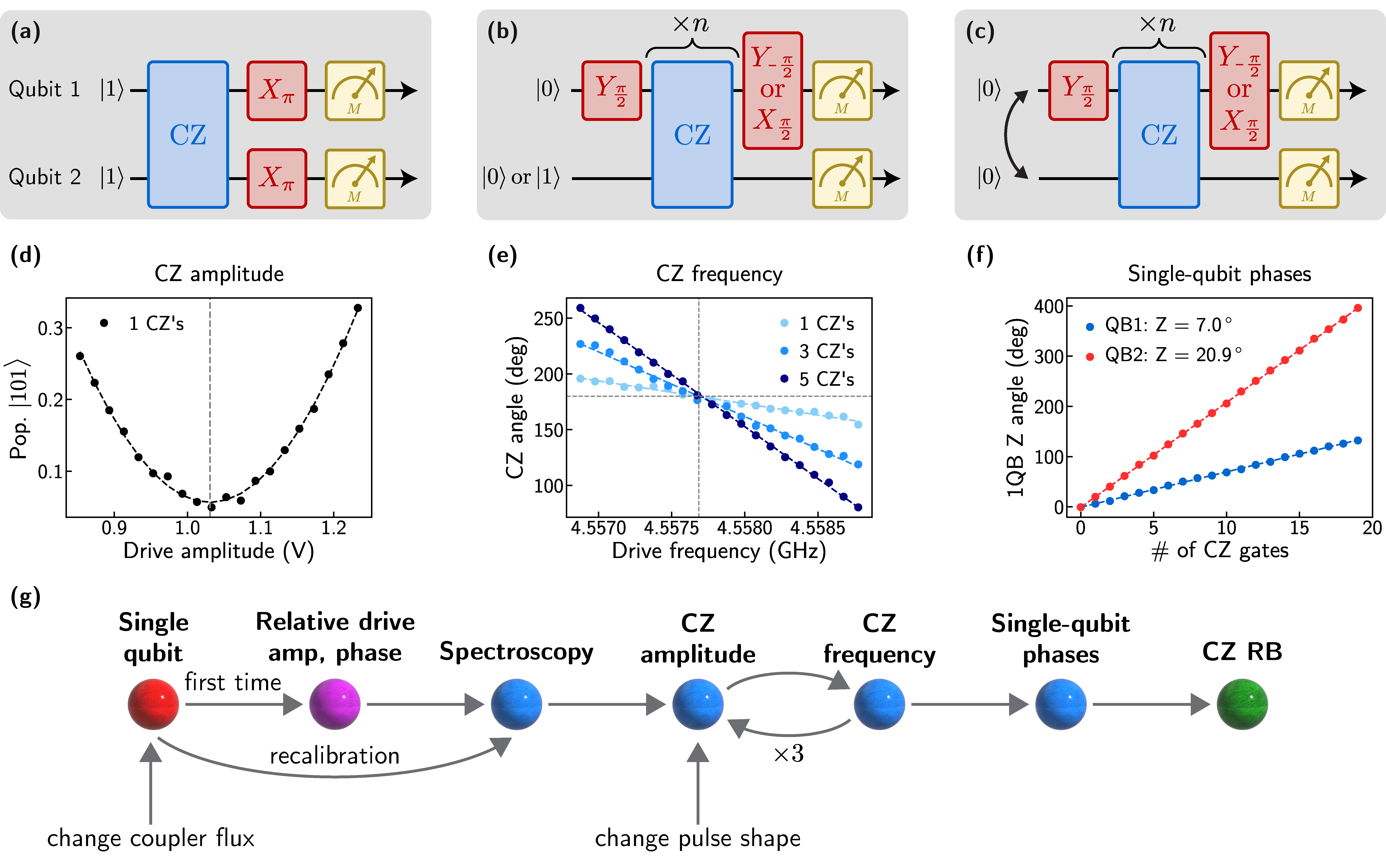}
\caption{\label{fig:sup_2qb_cal}
\textbf{CZ Gate calibration procedure.} \cpanel{a-c} Pulse sequences for \cpanel{d-f} respectively. The double arrow in (c) indicates that the same pulse sequence was performed twice, with qubits 1 and 2 exchanged. \cpanel{d} Calibrating the global amplitude of the CZ drive by minimizing the leakage. \cpanel{e} Calibrating the frequency of the CZ drive by measuring a conditional phase accumulation via Ramsey-like measurements. Each gate should contribute a~$180^\circ$ conditional phase shift. \cpanel{f} Measuring the single-qubit phase accumulations per CZ gate using the same Ramsey-like measurements. \cpanel{g} Graphical illustration of the full two-qubit calibration routine. When recalibrating the system for small flux drifts or periodic check-ins, we found it unnecessary to recalibrate the relative drive amplitude or phase.}
\end{figure*}

To benchmark our CZ gate, we performed two-qubit interleaved randomized benchmarking, in which Clifford gates were sampled from the two-qubit Clifford group, generated by~${\{I,\, \pm X,\, \pm Y,\, \pm X_{\pi/2},\, \pm Y_{\pi/2},\, \text{CZ}\}}$. Our specific decomposition results in 8.25 single-qubit gates and 1.5 CZ gates per Clifford. 

Interleaved randomized benchmarking consists of two steps: (1) performing standard randomized benchmarking, in which we extract the depolarizing parameter by measuring sequence fidelity with initial state~$\ket{00}$ and fitting to the decay~${p_{00} = A p^m + B}$. (2) Performing the same standard randomization benchmarking sequence, except an additional CZ gate is inserted between each pair of Cliffords, and the final recovery gate is modified accordingly. Fitting this decay gives~$p_\text{int}$, the interleaved depolarizing parameter. By comparing the relative decays between the two steps, the average CZ gate fidelity is extracted as
\begin{equation}
    F_\text{CZ} = 1 - (d - 1)(1 - p_\text{int}/p)/d,
\end{equation}
and its uncertainty is once again represented as a standard error, following the procedure of \aref{appendix:single_qubit}.

\section{Reinforcement learning} \label{appendix:RL}

As described in the main text, we used a model-free reinforcement learning agent to boost the gate fidelity beyond the physics-based calibration described in the previous section. 
Specifically, we used an algorithm known as proximal policy optimization (PPO). 

Application of reinforcement learning to quantum control has been discussed in detail in the literature. 
Here, we closely followed the procedure discussed in~\cite{Sivak2022} and implemented in~\cite{Sivak2023}, with code adapted from~\cite{Sivak2022git} and built on TF-Agents. As seen in \fref{fig:fidelities}\panel{b-c}, each round of training used 300 epochs and lasted roughly 75 minutes. We found the $\sim$15 second training time per epoch to be almost completely dominated by AWG waveform upload time, which took over 12 seconds for the data shown in \fref{fig:fidelities}. As updating the agent's policy according to PPO took comparably less time (less than 1 second), we elected to simply run PPO on the measurement computer instead of, for example, a separate computer equipped with a GPU. 
Since averaging was also relatively inexpensive compared to pulse upload time, we also elected to average the measurement results for each trial pulse 1000 times. 

\begin{table}[!htb]
\caption{\label{tab:RL}
\textbf{Reinforcement learning parameters.} Hyper-parameters of the PPO algorithm used to optimize the CZ gate. 
}
\begin{tabular}{|c|c|}
\hline
Learning rate & 0.01\\
\hline
Number of policy updates & 20\\
\hline
Importance ratio clipping & 0.1\\
\hline
Batch size & 30\\
\hline
Number of averages & 1000\\
\hline
  Value prediction loss coefficient   & 0.005\\
\hline
Gradient clipping & 1.0\\
\hline
Log probability clipping & 0.0\\
\hline
\end{tabular}
\end{table}

\section{CZ Gate vs. flux} \label{appendix:rb_vs_flux}

\begin{figure}[!htb]
\includegraphics{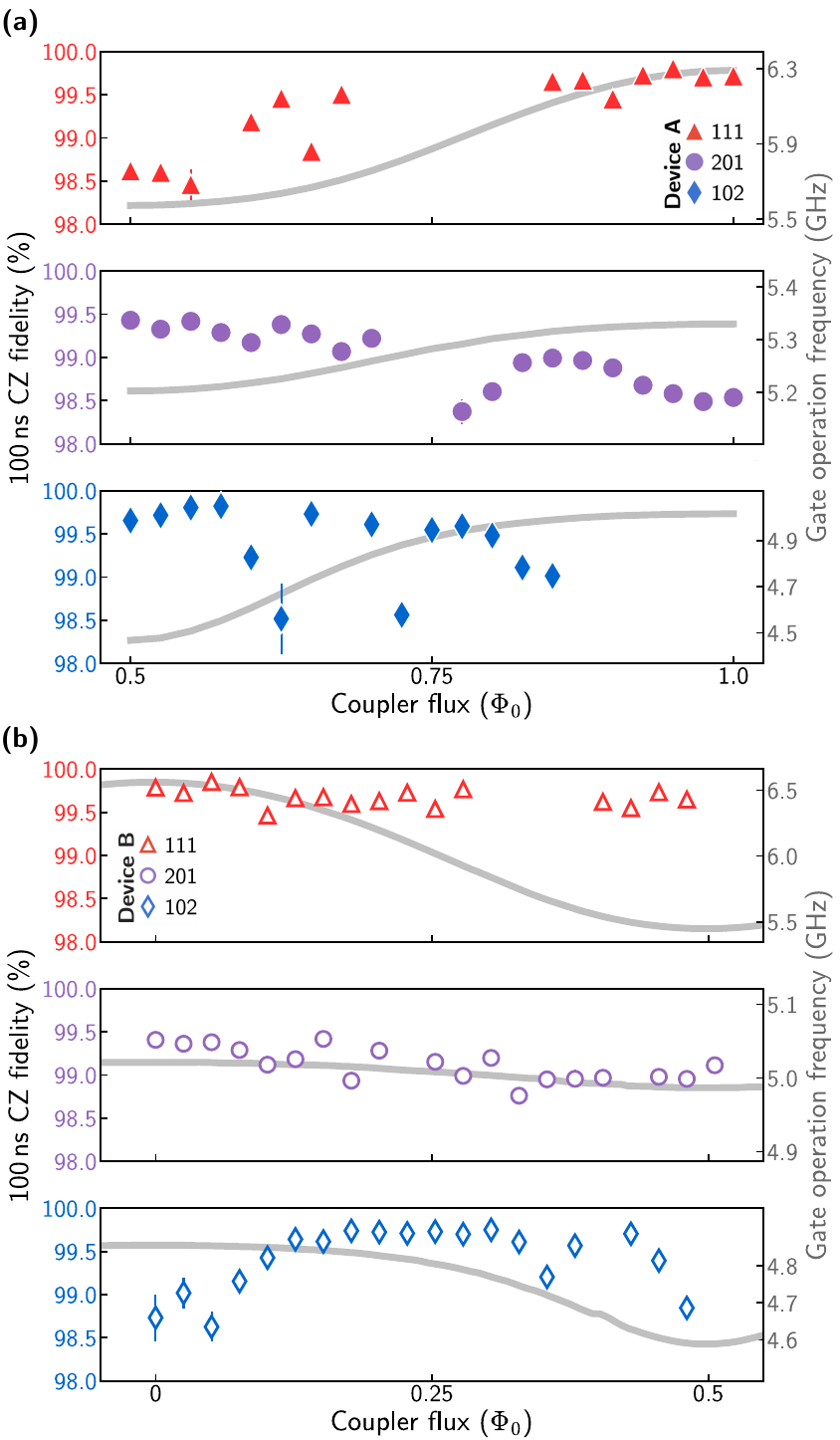}
\caption{\label{fig:sup_rb_vs_flux} \textbf{Alternate plotting of the data in \fref{fig:map_spec}\panel{c}.} Gate fidelities with a fixed \SI{100}{ns} pulse width as a function of the coupler flux for Device A \cpanel{a} and Device B \cpanel{b}. The drive frequency is plotted in gray, represented on the right axis. All points missing from the full set of 21 correspond to more severe calibration failures.
}
\end{figure}

\figureref{fig:sup_rb_vs_flux} shows an alternate plotting of the data in \fref{fig:map_spec}\panel{c}, illustrating more explicitly what frequencies each calibrated gate corresponds to.
As mentioned in the main text, each plot was created from a fully-automated measurement across 21 linearly spaced values of the coupler flux.
While many calibrated points have low fidelity, the most important aspect of this architecture is that there exist multiple transitions for which high fidelity is achievable, allowing flexibility in the operation frequency.
Nevertheless, we discuss the most common failure mechanisms and reasons for low fidelity to shed additional insights on the limits of our automated calibration versus the inherent limit of the transition being driven. 

The most common reason for low-performance operation points was nearby unwanted transitions, leading to a high amount of leakage. This is most evident in Device A, in which the~$\ket{101}\leftrightarrow\ket{111}$ transition at~$\Phi_\text{ext,c} = \SI{1.0}{\Phi_0}$, and the~$\ket{101}\leftrightarrow\ket{102}$ transition at~$\Phi_\text{ext,c} = \SI{0.5}{\Phi_0}$ are furthest detuned from their nearest unwanted transition, resulting in higher fidelities in these regions. Where these unwanted transitions have a much smaller detuning, we expect to obtain higher fidelities by increasing the drive pulse beyond \SI{100}{ns}. In other regions, we expect higher fidelities by decreasing the gate time.

A second common mechanism for failed calibration was the inability to find a drive frequency corresponding to a~$180^\circ$ conditional phase shift. Due to the large hybridization, we occasionally measured ac-Stark shifts large enough to counter-balance the natural change in the conditional-phase angle as a function of drive detuning. The calibration could be recovered by either using a slower gate or by adjusting the relative drive amplitudes to tweak the total ac-stark shift. 

Other less common mechanisms for a failed calibration include TLSs or accidental resonances with undesired transitions, including higher-photon transitions. 

\begin{figure}[!htb]
\includegraphics{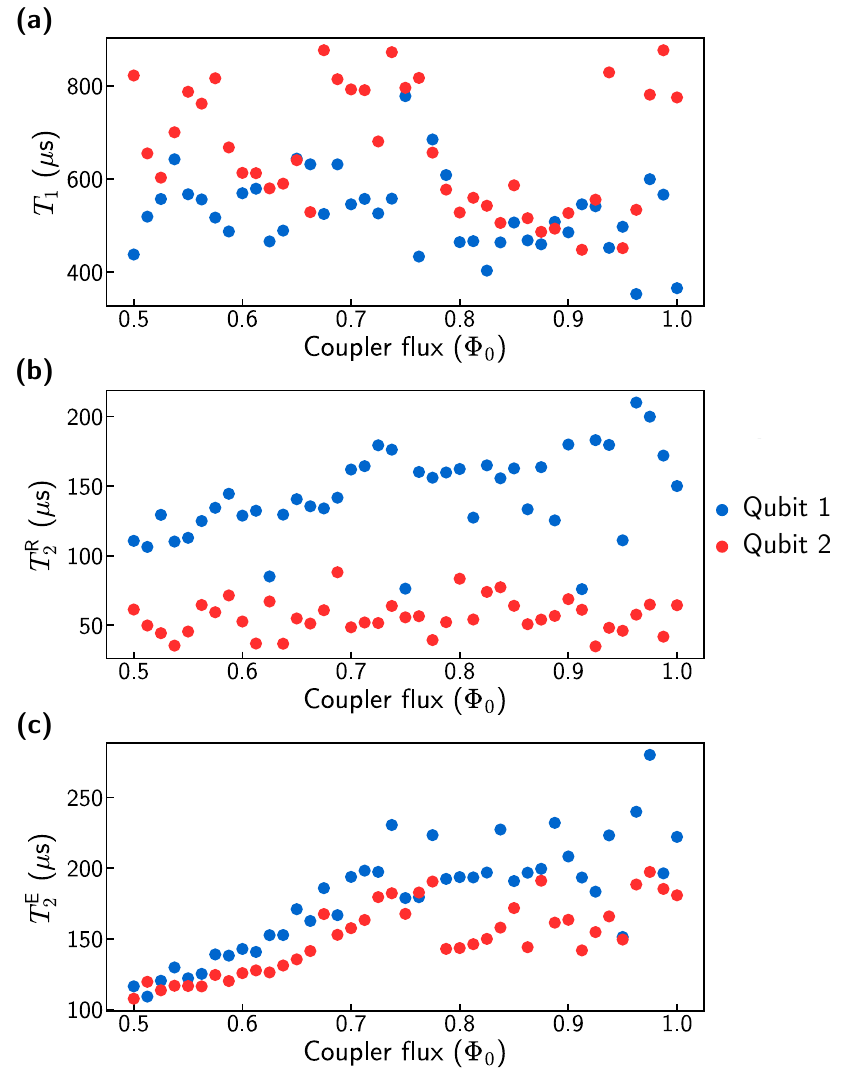}
\caption{\label{fig:sup_coherence_vs_flux_D1} \textbf{Qubit coherences in Device A with both qubits biased at \SI{0.5}{\Phi_0}}. Panels \cpanel{a}, \cpanel{b}, \cpanel{c} show the~$T_1$, Ramsey, and echo decay times respectively as a function of the coupler flux. All decays were fit to an exponential and measurements were performed over a \SI{12}{hour} period. 
}
\end{figure}

\figureref{fig:sup_coherence_vs_flux_D1} shows the individual qubit coherence times as a function of the local coupler flux in Device A. Each coherence time was simultaneously measured on each qubit, with both qubits precisely re-tuned to~$\Phi_\text{ext} = \SI{0.5}{\Phi_0}$ for each value of the coupler flux. Notably, these coherence times are slightly shorter than those listed in \tref{tab:hamiltonian}, due to bias-induced heating of the local flux lines. However, these coherence times are a more accurate representation of the quality of the qubits when performing two-qubit gates and simultaneous single-qubit gates.
When performing simultaneous single-qubit gates, we biased the coupler at roughly~$\Phi_\text{ext,c} = \SI{0.77}{\Phi_0}$, corresponding to no current being sent through the coupler flux line. We suspect the low Ramsey time of qubit 2 to be caused by Aharonov-Casher dephasing from coherent quantum phase slips~\cite{DiPaolo2021, Mizel2020}, based on its Hamiltonian parameters. 

\section{Error modeling} \label{appendix:error}
In this section, we build up an analytic error budget to estimate the impact of various types of coherent and incoherent errors on gate fidelities. We model our gates as a completely positive trace-preserving (in some subspace) map~$\mathcal{G}$ acting on an input state~$\rho$. The Kraus representation theorem then allows us to express all such processes as~$\mathcal{G}(\rho) = \sum_k G_k \rho G_k^\dagger$ for some set of Kraus operators~$G_k$ obeying the normalization condition~$\sum_k G_k^\dagger G_k = I$. The average state fidelity of such a process~$\mathcal{G}$ is then given by

\begin{equation} \label{eq:average_state_fidelity}
F = \frac{1}{n(n+1)} \left[\text{Tr}\left(\sum_k M_k^\dagger M_k\right) + \sum_k |\text{Tr}(M_k)|^2 \right],
\end{equation}
where~$M_k = P U_0^\dagger G_k P$, and~$n$ is the dimension of the computational subspace~\cite{Pedersen2007}. Also,~$P$ is the projection operator onto the computational subspace, and~$U_0$ is the ideal unitary operation of the process. We reproduce here the error corresponding to relaxation ($T_1$) and pure (Markovian) dephasing ($T_\phi$) corresponding to a gate of length~$t_g$.
\begin{equation}
    F_\text{1 qubit} = 1 - \frac{t_g}{3}\left(\frac{1}{T_1} + \frac{1}{T_\phi}\right)
\end{equation}

\begin{equation}
    F_\text{2 qubits} = 1 - \frac{4 t_g}{5}\left(\frac{1}{T_1} + \frac{1}{T_\phi}\right)
\end{equation}
One critical assumption in these well-known formulas is that the gate operation stays within the computational subspace, an invalid assumption for our two-qubit gate. 

\subsection{Relaxation of higher energy levels}
In our two-qubit gate, there are five relevant states: the computational states~$\{\ket{00}, \ket{01}, \ket{10}, \ket{11}\}$ and the non-computational state we drive to~$\ket{\alpha}$. For mathematical simplicity, we imagine modeling an identity gate composed of two CZ gates. The error per unit time will not change, and this allows us to use~$U_0 = I$ as well as simplifies the phases in our Kraus operators. We model the incoherent decay under driven evolution as if~$\ket{11}$ and~$\ket{\alpha}$ decay into each other at equivalent rates (a valid assumption when the gate time~$t_g$ is small compared to the relaxation rate~$T_{1,\alpha}$). Assuming no other decay channels, the full set of Kraus operators for this process is 

\begin{equation}
G_0 = \frac{1}{\sqrt{2}} \begin{pmatrix}
1 & 0 & 0 & 0 & 0\\
0 & 1 & 0 & 0 & 0\\
0 & 0 & 1 & 0 & 0\\
0 & 0 & 0 & 1 & 0\\
0 & 0 & 0 & 0 & e^{-t/2T_1}
\end{pmatrix}
\end{equation}

\begin{equation}
\hspace{1em} G_1 = \frac{1}{\sqrt{2}} \begin{pmatrix}
0 & 0 & 0 & 0 & 0\\
0 & 0 & 0 & 0 & 0\\
0 & 0 & 0 & 0 & 0\\
0 & 0 & 0 & 0 & \sqrt{1 - e^{-t/T_1}}\\
0 & 0 & 0 & 0 & 0
\end{pmatrix}
\end{equation}

\begin{equation}
G_2 = \frac{1}{\sqrt{2}} \begin{pmatrix}
1 & 0 & 0 & 0 & 0\\
0 & 1 & 0 & 0 & 0\\
0 & 0 & 1 & 0 & 0\\
0 & 0 & 0 & e^{-t/2T_1} & 0\\
0 & 0 & 0 & 0 & 1
\end{pmatrix}
\end{equation}

\begin{equation}
G_3 = \frac{1}{\sqrt{2}} 
\begin{pmatrix}
0 & 0 & 0 & 0 & 0\\
0 & 0 & 0 & 0 & 0\\
0 & 0 & 0 & 0 & 0\\
0 & 0 & 0 & 0 & 0\\
0 & 0 & 0 & \sqrt{1 - e^{-t/T_1}} & 0
\end{pmatrix}
\end{equation}
One can verify the behavior of these operators by computing that~${\bra{11}\mathcal{G}(\rho)\ket{11} = \frac{\rho_{11} + \rho_{\alpha}}{2} + \frac{\rho_{11} - \rho_\alpha}{2} e^{-t/T_{1,\alpha}}}$. Finally, we take the projection operator to be~$P = \ketbra{00} + \ketbra{01} + \ketbra{10} + \ketbra{11}$. Inserting these operators into \eref{eq:average_state_fidelity} and Taylor expanding in~$t_g / T_{1, \alpha} < 1$, we obtain
\begin{equation}
    F \approx 1 - \frac{1}{8} \frac{t_g}{T_{1,\alpha}}.
\end{equation}

\subsection{CZ phase error}
We consider an error in phase calibration, in which we successfully return all population back to the computational subspace, but with a~$\ket{11}$ state phase of~$\pi + d\phi$. In the unitary special case of \eref{eq:average_state_fidelity} ($\mathcal{G}(\rho) = U \rho U^\dagger$), we need only compute
\begin{align}
M &= U_0^\dagger U\\
&= 
\begin{bmatrix}
1 & 0 & 0 & 0\\
0 & 1 & 0 & 0\\
0 & 0 & 1 & 0\\
0 & 0 & 0 & -1
\end{bmatrix}
\begin{bmatrix}
1 & 0 & 0 & 0\\
0 & 1 & 0 & 0\\
0 & 0 & 1 & 0\\
0 & 0 & 0 & e^{-i \pi - i d\phi}
\end{bmatrix}\\
&= 
\begin{bmatrix}
1 & 0 & 0 & 0\\
0 & 1 & 0 & 0\\
0 & 0 & 1 & 0\\
0 & 0 & 0 & e^{- i d\phi}
\end{bmatrix}.
\end{align}
Inserting this into \eref{eq:average_state_fidelity}, we obtain
\begin{equation}
    F = \frac{7 + 3\cos(d\phi)}{10} \approx 1 - \frac{3}{20}d\phi^2.
\end{equation}
Values of note are that for a fidelity of 99.9\%, we can tolerate a phase error of~$4.7^\circ$ and for a fidelity of 99.99\%, we can tolerate a phase error of~$1.5^\circ$. 
We can similarly convert this into an error on drive frequency, assuming the drive frequency is the sole degree of freedom in tuning the aforementioned phase. The geometric phase accumulation associated with some frequency change~$\delta$ of a full-period driven oscillation is~$\delta t_g/2$. The angle errors above then translate into \SI{0.5}{MHz} error for 99.9\% fidelity and \SI{160}{kHz} for 99.99\% fidelity. 

\subsection{Amplitude error}
In calibrating the Rabi oscillation driving our CZ gate, we chose a fixed gate time and calibrated the amplitude of the pulse to obtain a single-period oscillation. The unitary corresponding to this Rabi rotation in our five-state Hilbert space~$\{\ket{00}, \ket{01}, \ket{10}, \ket{11} \ket{\alpha}\}$ is

\begin{equation}
U = \begin{pmatrix}
1 & 0 & 0 & 0 & 0\\
0 & 1 & 0 & 0 & 0\\
0 & 0 & 1 & 0 & 0\\
0 & 0 & 0 & \cos(\Omega t / 2) & -i \sin(\Omega t / 2)\\
0 & 0 & 0 & -i \sin(\Omega t / 2) & \cos(\Omega t / 2)\\
\end{pmatrix}
\end{equation}

\noindent where~$\Omega$ is the Rabi oscillation of the CZ pulse. Projecting onto the computational subspace ($P$) and assuming an ideal CZ unitary 
\begin{equation}
U_0 = \ketbra{00} + \ketbra{01} + \ketbra{10} - \ketbra{11},
\end{equation}
we insert~$M = P U_0^\dagger U P$ into~\eref{eq:average_state_fidelity} to obtain
\begin{equation}
    F = \frac{1}{10}(6 - 3\cos(\Omega t / 2) + \cos^2(\Omega t / 2)) \approx 1 - \frac{1}{16} d\theta^2.
\end{equation}
Converting this amplitude error into a phase error 
${d\theta = \Omega t - 2\pi}$, 99.9\% fidelity corresponds to a~$7.25^\circ$ error and 99.99\% fidelity corresponds to a~$2.29^\circ$ error. To relate this more directly to our experimental apparatus, assuming that the pulse is calibrated to a roughly \SI{1}{V} amplitude, these phases correspond to voltage errors of \SI{20}{mV} and \SI{6.4}{mV} respectively. 

We conclude this section by emphasizing that at the error levels discussed, calibration precision is quite lenient and that errors will be dominated by decoherence and other unmodeled behavior such as leakage through neighboring transitions. Furthermore, all calculations performed are meant to model the fidelity of a single-gate, which may not necessarily be equal to the fidelity extracted from interleaved randomized benchmarking due to the nature of coherent errors.  

\section{Device B} \label{appendix:other_device}

\begin{figure*}[!htb]
\includegraphics{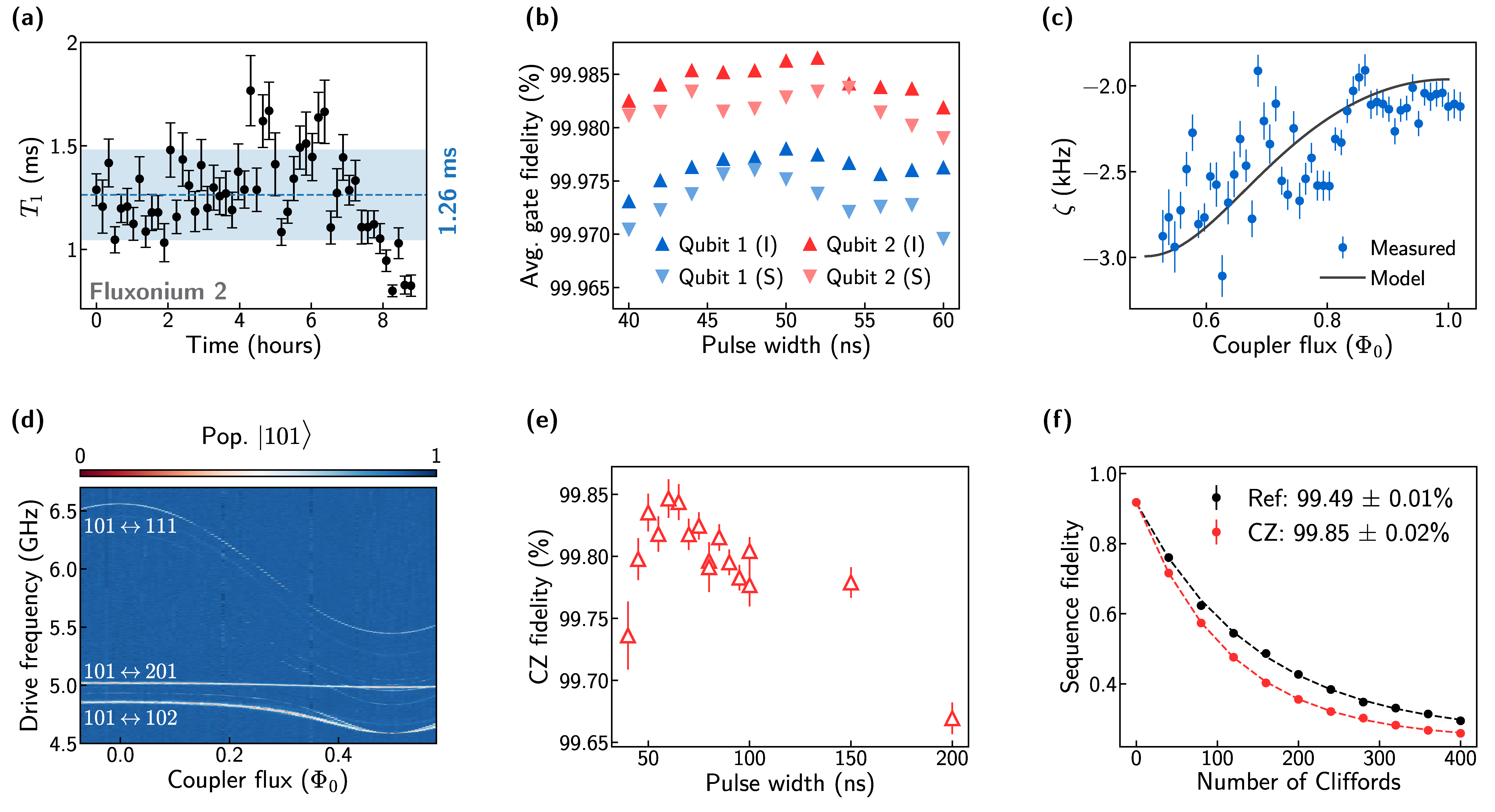}
\caption{\label{fig:sup_other_device} \textbf{Selected data from Device B.} \cpanel{a}  Repeated~$T_1$ measurements of fluxonium~2 over a roughly nine-hour time span. The dashed blue line indicates the median~$T_1$ of \SI{1.26}{ms}, and the blue shaded region encompasses~$\pm 1$ standard deviations.  \cpanel{b} Individual (I) and simultaneous (S) randomized benchmarking for both qubits, varying the width of a cosine pulse envelope. \cpanel{c} Measured~$ZZ$-interaction strength as a function of the coupler flux. \cpanel{d} Spectroscopy of the non-computational states which activate the two-qubit gate. \cpanel{e} CZ fidelities with varying pulse width, driving the~$\ket{101}\leftrightarrow\ket{111}$ transition at~$\Phi_\text{ext,c} = 0.063$. \cpanel{f} Reference and interleaved randomized benchmarking traces averaged over 20 randomizations for a pulse width of \SI{60}{ns}. 
}
\end{figure*}

In this section, we show data corresponding to Device B, a secondary device. Device B is an identically designed device, with extracted Hamiltonian parameters varying by up to 10\% as compared to Device A, typical of fabrication variations and differences in device aging. The coherence times and single-qubit gate fidelities remain consistently high across both devices, and in particular, fluxonium~2 on Device B exhibited a median~$T_1$ of \SI{1.26}{ms} averaged over \SI{9}{hours} [see \fref{fig:sup_other_device}\panel{a}]. This, along with the measured lifetimes of fluxonium~2 on Device A, points to a reliable process and design for achieving high lifetime qubits in a planar geometry. Curiously, the single-qubit gate fidelities in \fref{fig:sup_other_device}\panel{b} were found to be optimized near a pulse width of \SI{50}{ns}, a significant difference between the optimal pulse width of \SI{18}{ns} for Device A. We currently do not have an explanation for this discrepancy. 

We measured a nearly identical value (within \SI{1}{kHz}) of the~$ZZ$ interaction rate in this device [\fref{fig:sup_other_device}\panel{c}], supporting our claim that the~$ZZ$ reduction does not rely on any precise parameter matching and is a reliable method to achieve (absolute) values below \SI{10}{kHz}. Most importantly, despite changes of up to \SI{300}{MHz} in the~$\ket{1} \leftrightarrow \ket{2}$ transition frequencies of the fluxonium qubits, we could still demonstrate high-fidelity CZ gates across a large frequency range [see \fref{fig:map_spec}\panel{c} in the main text], with peak fidelities above 99.8\% [see \fsref{fig:sup_other_device}\panel{d-f}]. 

\clearpage
\bibliography{ftf_paper}

\end{document}